\documentclass[12pt]{article} %
\usepackage[utf8]{inputenc}
\usepackage[section]{placeins}
\usepackage[top=.6in, bottom=.8in, left=0.4in, right=0.4in]{geometry}
\usepackage{amsmath}
\numberwithin{equation}{section}
\usepackage{amssymb}
\usepackage{tikz}
\usepackage{ctable}
\usepackage{setspace}	
\usepackage{hyperref}
\usepackage{cleveref}
\usepackage{lmodern}
\usepackage{graphicx}
\usepackage{float}
\usepackage[margin=20pt, font=normal]{caption}
\usepackage[margin=20pt,font+=small,labelformat=parens,labelsep=space,skip=6pt,list=false,hypcap=false]{subcaption}
\usepackage{bbm}
\usepackage{xcolor}
\usepackage[bottom]{footmisc}
\usepackage{amsthm}
\usepackage{tabularx}
\usepackage{booktabs}
\usepackage{dcolumn}
\usepackage{newclude}
\usepackage{enumitem}
\usepackage{array}
\usepackage{comment}
\usepackage{indentfirst}
\usepackage{lscape}
\usepackage{pdflscape}
\usepackage[disable,textwidth=1.5in,textsize=tiny]{todonotes}
\usepackage{xr}

\usepackage[bibstyle=numeric, citestyle=authoryear, doi=false, url=true, backend=biber, maxbibnames=6, maxcitenames=4, uniquelist=false, uniquename=false, sorting=nyt]{biblatex}
\renewbibmacro{in:}{}
\newcommand\independent{\protect\mathpalette{\protect\independenT}{\perp}}
    \def\independenT#1#2{\mathrel{\setbox0\hbox{$#1#2$}%
    \copy0\kern-\wd0\mkern4mu\box0}} 
\newcommand{\indicator}[1]{\mathbbm{1}\{#1\}}
\let\d\relax
\newcommand{\d}[1]{\ \mathrm{d}{#1} }

\newcolumntype{R}{>{\raggedleft\arraybackslash}X}
\newcommand{\citet}{\textcite}
\newcommand{\subproof}[1]{\noindent \underline{\textit{#1}}}
  
\allowdisplaybreaks[1]

\crefname{assumption}{Assumption}{Assumptions}
\newtheorem{assumption}{Assumption}

\crefname{altassumption}{Alternative Assumption}{Alternative Assumptions}
\newtheorem{remark}{Remark}
\newtheorem{theorem}{Theorem}
\newtheorem{lemma}{Lemma}
\crefname{lemma}{Lemma}{Lemmas}
\newtheorem{proposition}{Proposition}

\newtheorem{example}{Example}

\newtheorem*{csa}{Copula Stability Assumption}
\makeatletter
\newenvironment{csa-assumption}
  {\csa\phantomsection\def\@currentlabel{Copula Stability Assumption}}
  {\endcsa}
\makeatother
\newtheorem*{ccsa}{Conditional Copula Stability Assumption}
\makeatletter
\newenvironment{ccsa-assumption}
  {\ccsa\phantomsection\def\@currentlabel{Conditional Copula Stability Assumption}}
  {\endccsa}
\makeatother

\DeclareMathOperator{\E}{\mathrm{E}}%

\DeclareMathOperator{\Cov}{\mathrm{Cov}}

\let\P\relax%
\DeclareMathOperator{\P}{\mathrm{P}}
\DeclareMathOperator{\F}{\mathrm{F}}
\DeclareMathOperator{\C}{\mathrm{C}}

\title{Bounds on Distributional Treatment Effect Parameters using Panel Data with an Application on Job Displacement\footnote{This is an updated version of my job market paper ``Job Displacement during the Great Recession:  Tight Bounds on Distributional Treatment Effect Parameters using Panel Data.''  I would like to thank Alper Arslan, Afrouz Azadikhah-Jahromi, Colin Cameron, Bill Collins, Scott Cunningham, Xu Cheng, Dakshina De Silva, Frank Diebold, Frank DiTraglia, Juan Carlos Escanciano, Dalia Ghanem, Collin Green, Federico Gutierrez, Jorgen Hansen, Atsushi Inoue, Catherine Maclean, Pedro Sant'Anna, Shu Shen, Artyom Shneyerov, Doug Webber, and seminar participants at Baylor University, the Bureau of Labor Statistics, Concordia University, the Consumer Financial Protection Bureau, Indiana University, Lancaster University, Mathematica Policy Research, Temple University, the University of California-Davis, the University of California-Merced, and the University of Pennsylvania for helpful comments.  I especially thank my advisor Tong Li for his detailed comments and support on this project.  Code for the methods developed in the paper is available in the \texttt{R} package \texttt{csabounds}.}}

\date{February 24, 2020}

\author{Brantly Callaway\thanks{Department of Economics, University of Mississippi, 319 Odom Hall, University, MS 38677.  Email: bmcallaw@olemiss.edu}}

\begin{document}

\maketitle

\abstract{\noindent This paper develops new techniques to bound distributional treatment effect parameters that depend on the joint distribution of potential outcomes -- an object not identified by standard identifying assumptions such as selection on observables or even when treatment is randomly assigned.  I show that panel data and an additional assumption on the dependence between untreated potential outcomes for the treated group over time (i) provide more identifying power for distributional treatment effect parameters than existing bounds and (ii) provide a more plausible set of conditions than existing methods that obtain point identification.  I apply these bounds to study heterogeneity in the effect of job displacement during the Great Recession.  Using standard techniques, I find that workers who were displaced during the Great Recession lost on average 34\% of their earnings relative to their counterfactual earnings had they not been displaced.  Using the methods developed in the current paper, I also show that the average effect masks substantial heterogeneity across workers.}

\vspace{50pt}


\vspace{40pt}


\pagebreak

\doublespace

\normalsize

\section{Introduction}

One of the key contributions of the modern treatment effects literature is to explicitly acknowledge that the effect of participating in a treatment can differ across different individuals, even individuals with identical observable characteristics (see, for example, \textcite{heckman-smith-clements-1997,heckman-2001, imbens-wooldridge-2009}).  Despite allowing for heterogeneous effects, most work in the treatment effects literature focuses on summary measures like the average treatment effect (ATE) or average treatment effect for the treated (ATT).  This paper considers bounds on the distribution of \textit{individual-level} treatment effects in the case where a researcher has access to panel data.  Individual-level treatment effect heterogeneity poses a particularly difficult challenge.  Unlike summary parameters such as the ATE, which only depend on the marginal distributions of treated and untreated ``potential'' outcomes, the distribution of the treatment effect depends on the \textit{joint} distribution of treated and untreated potential outcomes.  

The joint distribution of potential outcomes is not identified under common identifying assumptions such as selection on observables or even when individuals are randomly assigned to treatment.  In each of these cases, although the marginal distributions of treated and untreated potential outcomes are identified, the copula -- which ``couples'' the marginal distributions into the joint distribution and captures the dependence between the marginal distributions -- is not identified.  The fundamental reason why the joint distribution of treated and untreated potential outcomes is not identified is that, for each individual, either a treated potential outcome or an untreated potential outcome is observed (but not both).  

To give an example, suppose a researcher is interested in the fraction of workers who have higher earnings following job displacement than they would have had if they not been displaced.  Further, suppose hypothetically that workers are randomly assigned to being displaced or not being displaced.  In this case, the average effect of job displacement is identified -- it is given by the difference between average earnings of those who are randomly assigned to be displaced and those who are randomly assigned to not be displaced.  But the fraction of workers that benefit from displacement is not identified because, for workers randomly assigned to be displaced (non-displaced), where they \textit{would} be in the distribution of non-displaced (displaced) earnings is not known.  

Existing methods take two polar opposite approaches to identifying the joint distribution of potential outcomes.  One idea is to construct bounds on the joint distribution without imposing any assumptions on the unknown dependence (\citet{heckman-smith-clements-1997, fan-park-2009, fan-park-2010, fan-park-2012}).  In the case of job displacement, these bounds rule out that the effect of job displacement is the same across all individuals (which is an important finding), but they are less informative about other parameters of interest.  For example, the 10th percentile of the effect of job displacement (which is a measure of the effect of job displacement for those most negatively effected by job displacement) is bounded between 26\% lower earnings and 96\% lower earnings.  These bounds are also consistent with anywhere between 0\% and 81\% of workers having higher earnings following displacement than they would have had if they had not been displaced.  

Another approach is to assume that the dependence is known.  The leading choice is rank invariance.  This assumption says that individuals at a given rank in the distribution of treated potential outcomes would have the same rank in the distribution of untreated potential outcomes.  This is a very strong assumption.  In the case of job displacement, it imposes severe restrictions on how heterogeneous the effect of job displacement can be; for example, it prohibits any workers at the top of the distribution of non-displaced earnings from becoming unemployed or taking a part time job following displacement.  But the assumption is much stronger than that -- it prohibits displacement from even swapping the rank of any workers relative to their rank had they not been displaced.  

In light of (i) the implausibility of existing point-identifying assumptions and (ii) the wide bounds resulting from imposing no assumptions on the missing dependence, I develop new, tighter bounds on parameters that depend on the joint distribution of potential outcomes.  Unlike existing work which considers the case of cross-sectional data, I exploit having access to panel data.  Panel data presents a unique opportunity to observe, at least for some individuals, both their treated and untreated potential outcomes though these are observed at different points in time.  With panel data and under plausible identifying assumptions, the bounds on the joint distribution are much tighter -- in theory, the joint distribution could even be point identified.  

Even though panel data appears to be useful for identifying the joint distribution of potential outcomes, there are still some challenges.  Let $Y_{1t}$ denote treated potential outcomes in the last period, $Y_{0t}$ denote untreated potential outcomes in the last period, and $Y_{0t-1}$ denote untreated potential outcomes in the previous period.  Under the condition that no one is treated until the last period, then the joint distribution of $(Y_{1t},Y_{0t-1})$ is identified for the treated group -- this is the joint distribution of treated potential outcomes in the last period and untreated potential outcomes in the prior period which is observed for treated individuals.  Standard identifying assumptions may be used to identify the marginal distribution $Y_{0t}$ for the treated group.  But tighter bounds on the distribution of the treatment effect hinge on obtaining restrictions on the joint distribution of $(Y_{1t},Y_{0t})$ -- the joint distribution of treated and untreated potential outcomes in the last period.  Panel data alone does not provide these restrictions.  

The main assumption in the current paper is that the dependence (or copula) of untreated potential outcomes over time does not change over time.  I call this assumption the Copula Stability Assumption.  This assumption combined with the panel data setup mentioned above leads to identification of the joint distribution of $(Y_{0t}, Y_{0t-1})$ -- the joint distribution of untreated potential outcomes in the last two periods for the treated group.  With this joint distribution in hand, I utilize the following result: for three random variables, when two of the three bivariate joint distributions are known, then bounds on the third bivariate joint distribution are at least as tight as the bounds when only the marginal distributions are known (\citet{joe-1997}).  In the context of difference in differences models, previous work has used panel data to recover missing dependence in the current period from observed dependence in previous periods (\citet{callaway-li-2019}).  But that approach is not possible in the current context because the dependence between treated and untreated potential outcomes is never observed -- even in previous periods.  Instead panel data is informative about the dependence between untreated potential outcomes over time, which leads to bounds instead of point identification here.  
To utilize the Copula Stability Assumption requires that at least three periods of panel data are available.
In order to assess the validity of the \ref{ass:copula-stability}, I consider what additional conditions are required for the \ref{ass:copula-stability} to hold in models with time invariant unobserved heterogeneity and panel data.  I also show that the copula of earnings over time is stable over time and discuss how to ``pre-test'' the \ref{ass:copula-stability} in cases where a researcher has access to more than three periods of panel data.

The approach developed in the current paper is related to other work on tighter bounds on distributional treatment effect parameters.  \textcite{fan-park-2009,fan-park-2010,fan-guerre-zhu-2017,firpo-ridder-2019} bound parameters that depend on the joint distribution when covariates are available.  I discuss how this approach can be combined with the approach considered in the current paper to obtain even tighter bounds.  Another assumption that can bound parameters that depend on the joint distribution of potential outcomes is the assumption of Monotone Treatment Response (MTR) (\citet{manski-1997}).  \textcite{kim-2018} combines this assumption with the statistical bounds approach.  MTR would imply that earnings for displaced workers cannot be larger than earnings would have been had they not been displaced.  \textcite{frandsen-lefgren-2017} obtain tighter bounds on the joint distribution of potential outcomes by ruling out negative dependence between the potential outcomes.  \citet{masten-poirier-2020} construct breakdown frontiers which are informative about how robust results are to deviations from rank invariance assumptions.

There is also some empirical work studying the distributional effects of participating in a program.  \textcite{djebbari-smith-2008} use Fr\'{e}chet-Hoeffding bounds to study the distributional effects of the PROGRESA program in Mexico.  \textcite{carneiro-hansen-heckman-2003, abbring-heckman-2007}, among others, use factor models to identify the joint distribution of treated and untreated potential outcomes.  A common alternative approach to studying treatment effect heterogeneity is to see how the average treatment effect differs across groups which are defined by their observable characteristics (see the discussion in \textcite{bitler-gelbach-hoynes-2017} for more details about this approach and its relationship to treatment effect heterogeneity due to unobservables).  The current paper is also related to work on bounds under relatively weak assumptions in other contexts (for example, \textcite{manski-1990, manski-pepper-2000, blundell-gosling-ichimura-meghir-2007, kline-santos-2013, gechter-2016, kline-tartari-2016}).

I propose estimators of the bounds on the main parameters of interest.  These estimators depend on a number of first-step estimators of conditional distribution functions.  I propose estimating these conditional distribution functions using quantile regression and distribution regression (similar approaches are taken in \citet{melly-santangelo-2015,wuthrich-2019}).  I also provide the limiting distribution of the estimators of the bounds and propose using the numerical bootstrap (\citet{hong-li-2018}) to conduct inference.  These asymptotic results are related to work on inference on partially identified parameters that depend on the joint distribution of potential outcomes (\citet{fan-park-2010,fan-wu-2010}) and build on recent results on Hadamard directionally differentiable functions (\citet{fang-santos-2019,hong-li-2018,masten-poirier-2020}).  As an intermediate step, I develop some new results on distribution regression estimators with generated regressors.

I apply the methods developed in the paper to study heterogeneous effects of job displacement during the Great Recession.  Using standard techniques, I find that annual earnings of displaced workers were, on average, 34\% lower in 2011 than they would have been had the worker not been displaced.  Then, using the methods developed in the paper, I construct bounds on distributional treatment effect parameters that exploit having access to panel data.  These bounds are substantially tighter than existing bounds and provide a credible alternative to point identifying assumptions that are not likely to hold in the current application.  I estimate that the 10th percentile of earnings losses due to displacement is bounded between 46\% lower earnings and 89\% lower earnings.  I also find that at least 10\% of workers have higher earnings after displacement than they would have had if they had not been displaced. These findings indicate that there is substantial heterogeneity in the effect of job displacement, but they  would not be available using existing approaches.

\section{Parameters of Interest}

\paragraph{Notation} \ 

The notation used throughout the paper is very similar to the notation used in the treatment effects literature in statistics and econometrics.  All individuals in the population either participate or do not participate in a treatment.  Let $D=1$ for individuals that participate in the treatment and $D=0$ for individuals who do not participate in the treatment (to minimize notation, a subscript $i$ representing each individual is omitted throughout the paper except in a few cases to increase clarity).  The paper considers the case where panel data is available.  The baseline case considered in the paper is one where there are exactly three time periods though the results could be extended to the case with more time periods.  Throughout the paper, I use $s$ to represent a generic time period and $t$, $t-1$, and $t-2$ to represent particular time periods.  Each individual has potential outcomes in the treated and untreated states in each time period which are given by $Y_{1s}$ and $Y_{0s}$, respectively.  For each individual, only one of these potential outcomes is observed at each time period; I denote an individual's observed outcome in a particular time period by $Y_s$.  For individuals that are treated in period $s$, $Y_{1s}$ is observed, but $Y_{0s}$ is not observed.  For individuals that are untreated in period $s$, $Y_{0s}$ is observed but $Y_{1s}$ is unobserved.  
I make the following assumption
\begin{assumption} \label{ass:id1}
  The observed data consists of $n$ observations of $\{Y_{idt}, Y_{0it-1}, Y_{0it-2}, X_i, D_i\}$ which are independently and identically distributed.
\end{assumption}
\Cref{ass:id1} covers available data in the baseline case considered in the paper.  In particular, \Cref{ass:id1} says that the researcher observes outcomes in three periods.  The researcher may also observe a vector of covariates $X$ which, following much of the treatment effects literature (e.g., \citet{heckman-ichimura-todd-1997,abadie-2005}), I assume are time invariant.  \Cref{ass:id1} also says that individuals are first treated in the last period which implies that untreated potential outcomes are observed for both the treated group and the untreated group in periods $t-1$ and $t-2$.  That is,
\begin{align*}
  Y_t = DY_{1t} + (1-D)Y_{0t}, \ \ Y_{t-1} = Y_{0t-1}, \ \ \mathrm{and} \ \ Y_{t-2} = Y_{0t-2}
\end{align*}
\Cref{ass:id1} can be relaxed if additional periods are available or if treatment can occur in other periods besides the last one, but it represents a baseline case for tighter bounds and corresponds to the data used to study job displacement.

The next assumption is the starting point for the main identification results in the paper.\footnote{All of the results in the paper continue to go through after conditioning on covariates $X$.  Throughout most of this section, I omit conditioning on covariates to keep the notation simpler and focus on main ideas; however, bounds on parameters of interest can be tightened when there are available covariates by combining the results in the current paper with existing results on tightening bounds in the presence of covariates (\citet{fan-park-2009,fan-park-2010,fan-guerre-zhu-2017,firpo-ridder-2019}).  This is straightforward to do in practice (see the discussion in \Cref{rem:covariates} below).} 
\begin{assumption}\label{ass:id2}
  $\F_{Y_{1t}|D=1}$ and $\F_{Y_{0t}|D=1}$ are identified.
\end{assumption}

\Cref{ass:id2} says that the marginal distribution of treated potential outcomes for the treated group, $\F_{Y_{1t}|D=1}$, and the marginal distribution of untreated potential outcomes for the treated group, $\F_{Y_{0t}|D=1}$, are identified.  The first is not a strong assumption -- it is given by the distribution of observed outcomes for the treated group, $\F_{Y_t|D=1}$.  The second is a stronger assumption.  This counterfactual distribution would be identified if, for example, treatment were randomly assigned.  However, in cases with observational data, like job displacement, it requires some identifying assumption.  But there are many methods available to identify this counterfactual distribution.  
At any rate, the goal of the current paper is to go beyond the more standard objective of identifying this counterfactual distribution and learn about the joint distribution; thus, at this point, \Cref{ass:id2} considers the more standard problem of identifying the counterfactual marginal distribution to be solved.\footnote{\label{fn:idy0} There are some cases where the identifying assumption for $F_{Y_{0t}|D=1}$ may provide additional structure that could potentially tighten the bounds on parameters that depend on the joint distribution of treated and untreated potential outcomes.  See Footnote \ref{fn:ass2} for more discussion of this point.  I thank an anonymous referee for pointing this out.}

\Cref{ass:id2} implies that parameters that depend on the marginal distributions of treated and untreated potential outcomes for the treated group are identified.  These include the Average Treatment Effect on the Treated (ATT)\footnote{All the parameters mentioned in this section condition on being part of the treated group, but one may also be interested in these parameters for the entire population.  Panel data is most useful for identifying parameters conditional on being part of the treated group because only for the treated group does one observe both treated and untreated potential outcomes, albeit at different points in time.  Using the techniques presented in the current paper can still lead to bounds on parameters for the entire population by combining the bounds for the treated group presented in the current paper with bounds for the untreated group coming from existing statistical bounds.  These bounds will be tighter if a larger fraction of the population is treated.  I do not pursue bounds on parameters for the entire population throughout the rest of the paper.} 
\begin{align*}
  ATT = E[Y_{1t} - Y_{0t}|D=1]
\end{align*}
and the Quantile Treatment Effect on the Treated (QTT)
\begin{align*}
  QTT(\tau) = \F^{-1}_{Y_{1t}|D=1}(\tau) - \F^{-1}_{Y_{0t}|D=1}(\tau)
\end{align*}
for $\tau \in (0,1)$ and where $\F^{-1}_X(\tau) = \inf\{x : \F_X(x) \geq \tau\}$.
But distributional parameters that depend on the joint distribution of potential outcomes are not identified and these may be of considerable interest.  For job displacement, I focus primarily on the Distribution of the Treatment Effect for the Treated (DoTT) and closely related parameters that are simple functionals of the DoTT; \textcite{heckman-smith-clements-1997} and \textcite{firpo-ridder-2019} discuss other parameters in this class that may be of interest in other applications.  The DoTT is the fraction of individuals that experience a treatment effect less than some value $\delta$.  It is given by
\begin{align*}
	DoTT(\delta) = \P(Y_{1t} - Y_{0t} \leq \delta | D=1)
\end{align*}
One can estimate the DoTT for different values of $\delta$ and plot them.  An alternative approach, and the one that seems more useful for studying job displacement is to invert the DoTT to obtain the Quantile of the Treatment Effect on the Treated (QoTT) which is given by
\begin{align*}
  QoTT(\tau) = \inf\{\delta : DoTT(\delta) \geq \tau \}
\end{align*}
To give some examples, in the context of job displacement, $QoTT(0.05)$ is the 5th percentile of the individual level effect of job displacement -- these are the workers who experience some of the largest negative effects of job displacement.  $QoTT(0.5)$ is the median effect of job displacement.  And $QoTT(0.95)$ is the effect of job displacement for workers who have close to the highest earnings relative to what they would have if they had not been displaced.  %
Another interesting parameter for job displacement is the fraction of workers that have higher earnings following job displacement than they would have had if they had not been displaced.  This is given by $(1 - DoTT(0))$.  One can also consider the fraction of workers who are much worse off due to job displacement by considering $DoTT(\delta^*)$ for some large negative value $\delta^*$.

\subsection{The Identification Issue and Existing Solutions}

This section explains in greater detail the fundamental reason why the joint distribution of potential outcomes is not point identified except under strong assumptions.  First, by \Cref{ass:id2}, both the marginal distribution of treated potential outcomes for the treated group $\F_{Y_{1t}|D=1}$ and the marginal distribution of untreated potential outcomes for the treated group $\F_{Y_{0t}|D=1}$ are identified.  The first can be obtained directly from the data; the second is obtained under some identifying assumption which is assumed to be available.  \textcite{sklar-1959} demonstrates that joint distributions can be written as the copula function of marginal distributions in the following way
\begin{align} \label{eqn:sklars}
	\F_{Y_{1t}, Y_{0t} | D=1}(y_1, y_0) = C_{Y_{1t}, Y_{0t} | D=1}\left(\F_{Y_{1t}|D=1}(y_1), \F_{Y_{0t}|D=1}(y_0) \right)
\end{align}
where $C_{Y_{1t}, Y_{0t} | D=1}(\cdot, \cdot): [0,1]^2 \rightarrow [0,1]$.  This representation highlights the key piece of missing information under standard assumptions -- the copula function.  Using results from the statistics literature, one can still construct the so-called Fr\'{e}chet-Hoeffding bounds on the joint distribution (\citet{hoeffding-1940, frechet-1951}).  These bounds arise from considering two extreme cases: (i) when there is rank invariance between the two marginal distributions and (ii) when there is perfect negative dependence between the two distributions.  \textcite{heckman-smith-clements-1997} follow this procedure and find that it leads to very wide bounds in general.\footnote{In that paper and in the current paper, the Fr\'{e}chet-Hoeffding bounds can rule out the common effects model (i.e., that the effect of the treatment is the same for all individuals); however, these bounds are much less useful for understanding some other aspects of individual-level treatment effect heterogeneity.}  Moreover, that paper points out that under strong forms of negative dependence, the bounds do not seem to make sense in an application on the treatment effect of participating in a job training program.

At the other extreme, one could posit a guess for the copula.  In the cross-sectional case, the most common assumption is rank invariance between treated potential outcomes and untreated potential outcomes for the treated group (for clarity, in the current paper I refer to this assumption as \textit{cross-sectional rank invariance}).\footnote{Rank invariance is also sometimes called perfect positive dependence, comonotonicity, or rank permanence in the literature.  This assumption was first implicitly made in the earliest work on estimating the distributional effects of treatment (\citet{doksum-1974, lehmann-1974}) that compared the difference between treated quantiles and untreated quantiles and interpreted this difference as the treatment effect at that quantile.  There is also recent work on testing the assumption of rank invariance (\citet{bitler-gelbach-hoynes-2006, dong-shen-2018, frandsen-lefgren-2018})}  This type of rank invariance assumption says that, for individuals in the treated group, whatever their rank in distribution of the outcome is, they would have the same rank in the distribution of untreated potential outcomes.  When panel data is available, an alternative assumption is \textit{rank invariance over time}.  This assumption says that, for individuals in the treated group, if they did not participate in the treatment, they would maintain their same rank in the distribution of outcomes over time.  This assumption is strong enough to identify the joint distribution of potential outcomes.  Like the assumption of cross-sectional rank invariance, this assumption is very strong in many applications in economics.  For example, in the context of job displacement, it would require that, in the absence of job displacement, displaced individuals would have exactly the same rank in the earnings distribution as they did in the previous period.  I discuss these rank invariance assumptions in more detail in Supplementary Appendix \ref{app:additional-results}; see also \Cref{rem:ri} below.

\section{Identification}

In the previous section, I have argued that assumptions that directly replace the unknown copula in Equation \ref{eqn:sklars} are not likely to hold.  This section considers an alternative approach that does not substitute for the copula in Equation \ref{eqn:sklars} directly but limits the possibilities for the copula.  The next assumption is the main identifying assumption in the paper.

\begin{csa-assumption} \label{ass:copula-stability} 
	For all $(u,v) \in [0,1]^2$
	\begin{align*}
		\C_{Y_{0t}, Y_{0t-1}|D=1}(u,v) = \C_{Y_{0t-1}, Y_{0t-2}|D=1}(u,v)
	\end{align*}
\end{csa-assumption}

The \ref{ass:copula-stability} says that the dependence between untreated potential outcomes at periods $t$ and $t-1$ is the same as the dependence between untreated potential outcomes at periods $t-1$ and $t-2$.  This assumption is useful because the dependence between untreated potential outcomes at period $t$ and period $t-1$ is not observed. Although, by assumption,  the counterfactual distribution of untreated potential outcomes for the treated group, $\F_{Y_{0t}|D=1}$, is identified and the distribution of untreated potential outcomes for the treated at period $t-1$, $\F_{Y_{0t-1}|D=1}$, is identified because untreated potential outcomes are observed for the treated group at period $t-1$, their dependence is not identified because $Y_{0t}$ and $Y_{0t-1}$ are not simultaneously observed for the treated group.  The \ref{ass:copula-stability} recovers the missing dependence.  This implies that the joint distribution of untreated potential outcomes at times $t$ and $t-1$ for the treated group, $\F_{Y_{0t},Y_{0t-1}|D=1}$, is identified.  This joint distribution is not of primary interest in the current paper.  But knowledge of this joint distribution is important for deriving tighter bounds on the distributions and parameters of interest.

To better understand the \ref{ass:copula-stability}, it is helpful to consider some examples. As a first example, the \ref{ass:copula-stability} says that if untreated potential outcomes at period $t-1$ are independent (or rank invariant) of untreated potential outcomes at period $t-2$, then untreated potential outcomes at period $t$ will continue to be independent (or rank invariant) of untreated potential outcomes at period $t-1$.  %
Or, for example, suppose the copula for $(Y_{0t-1},Y_{0t-2})|D=1$ is Gaussian with parameter $\rho$, the \ref{ass:copula-stability} says that the copula for $(Y_{0t},Y_{0t-1})|D=1$ is also Gaussian with parameter $\rho$ though the marginal distributions of outcomes can change in unrestricted ways.  For example, the distribution of earnings can shift over time or could become more unequal over time.  Likewise, if the copula is Archimedean, the \ref{ass:copula-stability} says that the generator function does not change over time.  For Archimedean copulas with a scalar parameter having a one-to-one mapping to dependence parameters such as Kendall's Tau or Spearman's Rho (examples include common Archimedean copulas such as the Clayton, Frank, and Gumbel copulas), the \ref{ass:copula-stability} says that the dependence parameter is the same over time.\footnote{One could also make the Copula Stability Assumption conditional on some covariates $X$.  This type of assumption might be more plausible in some applications.  For example, earnings over time may be more strongly positively dependent for older workers than for younger workers.}  

\begin{assumption} (Outcomes are continuously distributed) \label{ass:id3}

$Y_{0t}$, $Y_{0t-1}$ and $Y_{0t-2}$ are continuously distributed conditional on $D=1$. 
\end{assumption}
\Cref{ass:id3} is helpful for utilizing the \ref{ass:copula-stability}.  Continuously distributed outcomes imply that $C_{Y_{0t-1},Y_{0t-2}|D=1}$ is uniquely identified from the sampling process.  In practice, this assumption allows for the quantile functions in the expressions below to be well-defined.  %
Importantly, \Cref{ass:id3} does not require that $Y_{1t}$ is continuously distributed.  In the application on job displacement, this allows for some individuals to not be employed following job displacement because it can allow for a mass point in the distribution of treated potential outcomes.  

Next, as a preliminary result, I show that, under the \ref{ass:copula-stability}, the joint distribution of $(Y_{0t},Y_{0t-1})|D=1$ is identified.  Recall that this is not the joint distribution of interest, but identifying this joint distribution is going to provide identifying power for distributional treatment effect parameters that depend on the joint distribution of $(Y_{1t},Y_{0t})|D=1$.  
\begin{lemma}  \label{lem:idy0} Under \Cref{ass:id1,ass:id2,ass:id3} and the \ref{ass:copula-stability},
  \begin{align*}
    F_{Y_{0t},Y_{0t-1}|D=1}(y_0, y') = F_{Y_{0t-1},Y_{0t-2}|D=1}\Big(F^{-1}_{Y_{0t-1}|D=1} \circ F_{Y_{0t}|D=1}(y_0), F^{-1}_{Y_{0t-2}|D=1} \circ F_{Y_{0t-1}|D=1}(y')\Big)
  \end{align*}
  and
  \begin{align*}
    F_{Y_{0t}|Y_{0t-1},D=1}(y_0|y') &= F_{Y_{0t-1}|Y_{0t-2},D=1}\Big(F^{-1}_{Y_{0t-1}|D=1} \circ F_{Y_{0t}|D=1}(y_0) \Big| F^{-1}_{Y_{0t-2}|D=1} \circ F_{Y_{0t-1}|D=1}(y')\Big) \\
                                    &= P\Big(Y_{0t-1} \leq F^{-1}_{Y_{0t-1}|D=1} \circ F_{Y_{0t}|D=1}(y_0) \Big| F^{-1}_{Y_{0t-1}|D=1} \circ F_{Y_{0t-2}|D=1}(Y_{0t-2}) = y', D=1 \Big)
  \end{align*}
\end{lemma}

The proof of \Cref{lem:idy0} is provided in Supplementary Appendix \ref{app:model-proofs}.  The first part of Lemma \ref{lem:idy0} shows that the joint distribution of $(Y_{0t},Y_{0t-1})|D=1$ is identified under the \ref{ass:copula-stability} and gives an expression for it.  The second part provides an expression for the conditional distribution $F_{Y_{0t}|Y_{0t-1},D=1}$ which turns out to be useful later.  The intuition for \Cref{lem:idy0} is that: 
the \ref{ass:copula-stability} implies that one can learn about the joint distribution $(Y_{0t},Y_{0t-1})|D=1$ from the joint distribution $(Y_{0t-1},Y_{0t-2})|D=1$, but changes in the marginal distributions over time are unrestricted and therefore need to be adjusted.  This is what the terms like $F^{-1}_{Y_{0t-1}|D=1}\circ F_{Y_{0t}|D=1}(y_0)$ do; they take the distribution of untreated potential outcomes at time period $t$ and adjust it back to the distribution of untreated potential outcomes in time period $t-1$.

Next, I show how the \ref{ass:copula-stability} can be used to derive tighter bounds on the joint distribution of potential outcomes.  The next result is a simple application of Fr\'{e}chet-Hoeffding bounds to a conditional distribution; it provides an important building block for constructing tighter bounds on the joint distribution of potential outcomes.

\begin{lemma} \label{lemma:joint-cond}  Under \Cref{ass:id1,ass:id2,ass:id3} and the \ref{ass:copula-stability}, bounds on the joint distribution of treated and untreated potential outcomes for the treated group conditional on outcomes in the previous period are given by
	\begin{align*}
		\F^L_{Y_{1t},Y_{0t}|Y_{0t-1},D=1}(y_1, y_0|y') \leq \F_{Y_{1t},Y_{0t}|Y_{0t-1},D=1}(y_1, y_0|y') \leq \F^U_{Y_{1t},Y_{0t}|Y_{0t-1},D=1}(y_1, y_0|y')
	\end{align*}
	where
	\begin{align*}
		\F^L_{Y_{1t},Y_{0t}|Y_{0t-1},D=1}(y_1, y_0|y') &= \max\{\F_{Y_{1t}|Y_{0t-1},D=1}(y_1|y') + \F_{Y_{0t}|Y_{0t-1},D=1}(y_0|y')-1,0\} \\
          \F^U_{Y_{1t},Y_{0t}|Y_{0t-1},D=1}(y_1, y_0|y') &= \min\{\F_{Y_{1t}|Y_{0t-1},D=1}(y_1|y') , \F_{Y_{0t}|Y_{0t-1},D=1}(y_0|y')\}
	\end{align*}
\end{lemma}
The next theorem is the main result for bounds on the joint distribution of potential outcomes for the treated group.

\begin{theorem} \label{theorem:joint} Under \Cref{ass:id1,ass:id2,ass:id3} and the \ref{ass:copula-stability}, bounds on the joint distribution of treated and untreated potential outcomes for the treated group are given by
	\begin{align*}
		\F^L_{Y_{1t},Y_{0t}|D=1}(y_1, y_0) \leq \F_{Y_{1t},Y_{0t}|D=1}(y_1, y_0) \leq \F^U_{Y_{1t},Y_{0t}|D=1}(y_1, y_0)
	\end{align*}	
	where
        \begin{align*}
		\F^L_{Y_{1t},Y_{0t}|D=1}(y_1, y_0) &= \E[\F^L_{Y_{1t},Y_{0t}|Y_{0t-1},D=1}(y_1, y_0|Y_{0t-1}) | D=1]\\
		\F^U_{Y_{1t},Y_{0t}|D=1}(y_1, y_0) &= \E[\F^U_{Y_{1t},Y_{0t}|Y_{0t-1},D=1}(y_1, y_0|Y_{0t-1}) | D=1]
	\end{align*}
        where $F^L_{Y_{1t},Y_{0t}|Y_{0t-1},D=1}(y_1,y_0|y')$ and $F^U_{Y_{1t},Y_{0t}|Y_{0t-1},D=1}(y_1,y_0|y')$ are given in \Cref{lemma:joint-cond}.
\end{theorem}

The bounds in \autoref{theorem:joint} warrant some more discussion.  First, these bounds will be tighter than the bounds without using panel data unless $Y_{0t-1}$ is independent of $Y_{1t}$ and $Y_{0t}$.  But in most applications in economics $Y_{0t}$ and $Y_{0t-1}$ are likely to be positively dependent.  On the other hand, the joint distribution will be point identified if either (i) $Y_{1t}$ and $Y_{0t-1}$ are perfectly positively dependent or (ii) $Y_{0t}$ and $Y_{0t-1}$ are perfectly positively dependent.  Item (i) is very similar to the assumption of rank invariance across treated and untreated groups (though it also includes a time dimension); Item (ii) is exactly the condition of rank invariance in untreated potential outcomes over time used as a point identifying assumption.  Together, these conditions imply that if either one of two natural limiting conditions hold, then the joint distribution of potential outcomes will be point identified.  Moreover, intuitively the bounds will be tighter in cases that are ``closer'' to either of these two limiting cases.  This means that even in the case where the limiting conditions do not hold exactly, one is still able to (substantially) tighten the bounds that would arise in the case without panel data.  I provide the intuition for this point in the next example and provide a more formal proof in the proposition that follows.

\begin{example} \label{ex:spearmans-rho}
Spearman's Rho is the correlation of the ranks of two random variables; i.e., $\rho_S = Corr(\F_1(X_1), \F_2(X_2))$.  Bounds on Spearman's Rho can be derived when two out of three joint distributions and all marginal distributions (exactly the case in the current paper) are known (\citet[][Theorem 8.19]{joe-2015}).  Because the ranks $\F_{Y_{1t}|D=1}(Y_{1t})$, $\F_{Y_{0t}|D=1}(Y_{0t})$, and $\F_{Y_{0t-1}|D=1}(Y_{0t-1})$ are uniformly distributed conditional on $D=1$, their covariance matrix is given by
\begin{align*}
	\Cov\left(\F_{Y_{1t}|D=1}(Y_{1t}), \F_{Y_{0t}|D=1}(Y_{0t}), \F_{Y_{0t-1}|D=1}(Y_{0t-1}) | D=1 \right) = \begin{pmatrix} 1 & \rho_{12} & \rho_{13} \\ \rho_{12} & 1 & \rho_{23} \\ \rho_{13} & \rho_{23} & 1 \end{pmatrix}
\end{align*}
Consider the case where $\rho_{13}$ and $\rho_{23}$ are identified and $\rho_{12}$ is not known.  $\rho_{12}$ is partially identified because the covariance matrix must be positive semi-definite.  This results in the condition that
\begin{align*}
	\rho_{13} \rho_{23} - \sqrt{\rho_{13}^2 \rho_{23}^2 + (1-\rho_{13}^2 - \rho_{23}^2)} \leq \rho_{12} \leq \rho_{13} \rho_{23} + \sqrt{\rho_{13}^2 \rho_{23}^2 + (1-\rho_{13}^2 - \rho_{23}^2)}
\end{align*}
The width of the bounds is $2 \sqrt{\rho_{13}^2 \rho_{23}^2 + (1-\rho_{13}^2 - \rho_{23}^2)}$, and it is easy to show that for fixed $\rho_{23}$ with $|\rho_{23}|<1$, the width of the bounds on $\rho_{12}$ is decreasing as $\rho_{13}$ increases for $\rho_{13}>0$, and the width of the bounds is decreasing as $\rho_{13}$ decreases for $\rho_{13}<0$.  When either $\rho_{13}$ or $\rho_{23}$ is equal to one in absolute value, $\rho_{12}$ is point identified.  This corresponds exactly to the case of rank invariance (or perfect negative dependence) mentioned above for point identification.  The intuition of this result is that as the copula moves ``closer'' to rank invariance or perfect negative dependence, the bounds on the joint distribution of interest shrink.
\end{example}

\begin{proposition} \label{prop:1} Fix the marginal distributions $F_{Y_{1t}|D=1}$, $F_{Y_{0t}|D=1}$, and $F_{Y_{0t-1}|D=1}$ and the conditional distribution $F_{Y_{1t}|Y_{0t-1},D=1}$.  Now consider two possibilities for $F_{Y_{0t}|Y_{0t-1},D=1}$ given by $F_1$ and $F_2$.  Assume that $F_1$, $F_2$, and $F_{Y_{1t}|Y_{0t-1},D=1}$ are stochastically increasing\footnote{For two random variables $X$ and $W$, their conditional distribution $F_{X|W}$ is said to be stochastically increasing if $1-F_{X|W}(x|w)$ is increasing in $w$ for all $x$.  For two conditional distributions $F_{X|W}$ and $G_{X|W}$, having the same marginal distributions of $X$ and $W$, $F_{X|W}$ is said to be more stochastically increasing than $G_{X|W}$ if $F^{-1}_{X|W}(G_{X|W}(x|w)|w)$ is increasing in $w$ for all $x$.  Stochastically increasing is a well-known dependence property and more stochastically increasing is a common dependence ordering (see \citet{yanagimoto-okamoto-1969, schriever-1987} as well as related discussion in \textcite{joe-1997, nelsen-2007}).} and that $F_{Y_{1t}|Y_{0t-1},D=1} \prec^{SI} F_1 \prec^{SI} F_2$ where $F \prec^{SI} G$ indicates that $G$ is more stochastically increasing than $F$.  Then, the bounds on the joint distribution given in Theorem \ref{theorem:joint} are at least as tight when $F_{Y_{0t}|Y_{0t-1},D=1} = F_2$ as when $F_{Y_{0t}|Y_{0t-1},D=1} = F_1$.
\end{proposition}

Proposition \ref{prop:1} is a key result in the paper.  It says that the bounds in the paper get tighter when there is stronger dependence between $Y_{0t}$ and $Y_{0t-1}$ (under the \ref{ass:copula-stability}, this will be true when the dependence between $Y_{0t-1}$ and $Y_{0t-2}$ is stronger).  An analogous result also holds for $Y_{1t}$ and $Y_{0t-1}$ -- if the dependence is strong, the bounds will be tight.  In the literature, assumptions of rank invariance have been made as approximations because in many applications there is strong positive dependence though less than rank invariance.  Proposition \ref{prop:1} implies that the results in the current paper will be valid when rank invariance assumptions are violated, but the bounds will be ``tight'' in the case where these assumptions are not too far from the truth.  This is likely to be the most relevant case in many applications.  When the bounds are applied to job displacement later, I show that there is strong positive dependence (both between $Y_{1t}$ and $Y_{0t-1}$ and between $Y_{0t-1}$ and $Y_{0t-2}$ for the treated group) but less than rank invariance.  This implies that the assumptions of rank invariance will be violated, but it also implies that the bounds can be tightened substantially over bounds that only use the information from the marginal distributions of $Y_{1t}$ and $Y_{0t}$, respectively.

\Cref{prop:1} also implies that bounds obtained under the \ref{ass:copula-stability} are robust to some violations of the \ref{ass:copula-stability}.  In particular, when there is stronger dependence (in terms of ``more stochastically increasing'') between $Y_{0t}$ and $Y_{0t-1}$ than there was for $Y_{0t-1}$ and $Y_{0t-2}$ for individuals in the treated group, then the bounds developed under the \ref{ass:copula-stability} are conservative.

Just as knowledge of $\F_{Y_{1t},Y_{0t-1}|D=1}$ and $\F_{Y_{0t},Y_{0t-1}|D=1}$ leads to bounds on the joint distribution of interest $\F_{Y_{1t},Y_{0t}|D=1}$, knowledge of these distributions can also be used to bound the DoTT, the QoTT, and other parameters that depend on the joint distribution.  These results are presented next.  Sharp bounds on the distribution of the treatment effect are known in the case where there is no additional information besides the marginal distributions (\citet{fan-park-2010}).  These bounds are obtained using results from the statistics literature for the distribution of the difference of two random variables when the marginal distributions are fixed (\citet{makarov-1982, ruschendorf-1982, frank-nelsen-schweizer-1987, williamson-downs-1990}).  I use these same bounds for the conditional distribution of the treatment effect.

\begin{lemma} \label{lemma:cond-dte} (Conditional Distribution of the Treatment Effect)	Under \Cref{ass:id1,ass:id2,ass:id3} and the \ref{ass:copula-stability}, bounds on the distribution of the treatment effect for the treated group conditional on the outcome in the previous period are given by
	\begin{align*}
		\F^L_{Y_{1t}-Y_{0t}|Y_{0t-1},D=1}(\delta|y') \leq \F_{Y_{1t}-Y_{0t}|Y_{0t-1},D=1}(\delta|y') \leq \F^U_{Y_{1t}-Y_{0t}|Y_{0t-1},D=1}(\delta|y')
	\end{align*}
	where
	\begin{align*}
		\F^L_{Y_{1t}-Y_{0t}|Y_{0t-1}, D=1}(\delta | y') &= \sup_y \max\{\F_{Y_{1t}|Y_{0t-1},D=1}(y|y') - \F_{Y_{0t}|Y_{0t-1},D=1}(y-\delta|y'), 0 \} \\
		\F^U_{Y_{1t}-Y_{0t}|Y_{0t-1}, D=1}(\delta | y') &= 1+\inf_y \min\{\F_{Y_{1t}|Y_{0t-1},D=1}(y|y') - \F_{Y_{0t}|Y_{0t-1},D=1}(y-\delta|y'), 0 \} 
	\end{align*}
\end{lemma}

The next result provides bounds for the DoTT.

\begin{theorem} (Distribution of the Treatment Effect) \label{theorem:dte} Under \Cref{ass:id1,ass:id2,ass:id3} and the \ref{ass:copula-stability}, bounds on $DoTT(\delta)$ are given by
  \begin{align*}
    DoTT^L(\delta) \leq DoTT(\delta) \leq DoTT^U(\delta)
  \end{align*}
  where
  \begin{align*}
    DoTT^L(\delta) = \F^L_{Y_{1t}-Y_{0t}|D=1}(\delta) &= \E[\F^L_{Y_{1t}-Y_{0t}|Y_{0t-1}, D=1}(\delta | Y_{0t-1}) | D=1] \\
    DoTT^U(\delta) = \F^U_{Y_{1t}-Y_{0t}|D=1}(\delta) &= \E[\F^U_{Y_{1t}-Y_{0t}|Y_{0t-1}, D=1}(\delta | Y_{0t-1}) | D=1]
	\end{align*}
	where $\F^L_{Y_{1t}-Y_{0t}|Y_{0t-1}, D=1}(\delta | y')$ and $\F^U_{Y_{1t}-Y_{0t}|Y_{0t-1}, D=1}(\delta | y')$ are given in \autoref{lemma:cond-dte}. 
\end{theorem}

\begin{remark}
  The results of Proposition \ref{prop:1} also imply that the bounds on the DoTT are tighter under the conditions given in Proposition \ref{prop:1}.  To be specific, consider two possibilities for $F_{Y_{0t}|Y_{0t-1},D=1}$ given by $F_1$ and $F_2$ with $F_1 \prec^{SI} F_2$ as in Proposition \ref{prop:1} and assume all other conditions as in that proposition.  Proposition \ref{prop:1} implies $C^L_1 \prec^{C} C^L_2$ (i.e., $C^L_2$ is more concordant than $C^L_1$) where $C^L_j$ denotes the lower bound on the copula of $Y_{1t}$ and $Y_{0t}$ when $F_{Y_{0t}|Y_{0t-1},D=1}=F_j$ for $j=1,2$.  Bounds on the DoTT get tighter when the lower bound of the copula is more concordant (\citet{williamson-downs-1990}).  This implies that the bounds on the DoTT will be tighter when there is more positive dependence between $Y_{0t}$ and $Y_{0t-1}$ (which occurs under the \ref{ass:copula-stability} when there is more positive dependence between $Y_{0t-1}$ and $Y_{0t-2}$).  
\end{remark}

\begin{remark} \label{rem:covariates}
  As mentioned above, the main identification results continue to go through when the assumptions hold conditional on covariates.  In particular, in this case, one can derive  bounds on the conditional DoTT, as in \Cref{lemma:cond-dte}, that are conditional on both $Y_{0t-1}$ and $X$.  Then, following the same arguments for the unconditional DoTT as in \Cref{theorem:dte}, one can derive bounds on the unconditional DoTT by averaging over both $Y_{0t-1}$ and $X$.  These bounds will be tighter than the bounds that do not include covariates which follows using the same arguments as in \citet{fan-park-2009,fan-park-2010,fan-guerre-zhu-2017,firpo-ridder-2019}.
\end{remark}

Bounds on the QoTT can be obtained from the bounds on the DoTT.  The upper bound on the QoTT comes from inverting the lower bound of the DoTT, and the lower bound on the QoTT comes from inverting the upper bound on the DoTT.

\begin{theorem} (Quantile of the Treatment Effect) \label{theorem:qotet}  Under \Cref{ass:id1,ass:id2,ass:id3} and the \ref{ass:copula-stability}, bounds on $QoTT(\tau)$ are given by
	\begin{align*}
		QoTT^L(\tau) \leq QoTT(\tau) \leq QoTT^U(\tau)
	\end{align*}
	where 
	\begin{align*}
		QoTT^L(\tau) &= \inf\{\delta : DoTT^U(\delta) \geq \tau\} \\
		QoTT^U(\tau) &= \inf\{\delta : DoTT^L(\delta) \geq \tau\} 
	\end{align*}
	and $DoTT^L(\delta)$ and $DoTT^U(\delta)$ are given in \autoref{theorem:dte}.
\end{theorem}

\section{How Plausible is the Copula Stability Assumption?}

Since the \ref{ass:copula-stability} is the key identifying assumption used in the paper and is crucial for exploiting panel data to deliver tighter bounds on the distributional treatment effect parameters considered in this paper, it is worth considering how plausible this assumption is.  In this section, I consider several types of evidence.  First, I consider what additional restrictions need to be satisfied in typical panel data models (particularly models with individual heterogeneity) in order for the \ref{ass:copula-stability} to hold.  Second, I propose two ``pre-tests'' for the \ref{ass:copula-stability} that are available in applications where there are more than two pre-treatment time periods.  Finally, Supplementary Appendix \ref{app:additional-results} considers in more detail the interpretation of the \ref{ass:copula-stability} in the case where the outcome of interest is an individual's earnings and shows that the copula of annual earnings over time has been remarkably stable in the United States since the mid 20th century.  

\subsection{Additional Conditions for the Copula Stability Assumption to Hold}

In this section, I consider two leading models for untreated potential outcomes when there is time invariant unobserved heterogeneity and when panel data is available: (i) two-way fixed effects models and (ii) the Change in Changes model (\citet{athey-imbens-2006}).

\subsubsection{Two-way Fixed Effects}

Consider the following two-way fixed effects model for untreated potential outcomes.  For $s \in \{ t, t-1, t-2\}$
\begin{align} \label{eqn:twfe}
  Y_{0is} = \theta_s + \eta_i + V_{is}
\end{align}
where $\theta_s$ is a time fixed effect, $\eta_i$ is time invariant unobserved heterogeneity that can be distributed differently for individuals in the treated group and untreated group, and $V_{is}$ are time varying unobservables.  \Cref{eqn:twfe} is a leading model in the treatment effects literature in the case with time invariant unobserved heterogeneity and panel data; in particular, it corresponds to the sort of model required for Difference in Differences designs to identify the ATT (see discussion in \citet{blundell-dias-2009}).  The next result provides conditions under which the \ref{ass:copula-stability} holds in the model in \Cref{eqn:twfe}.

\begin{proposition} \label{prop:twfe}
  In the two-way fixed effects model for untreated potential outcomes given above, and under the additional condition that $C_{\eta + V_t, \eta + V_{t-1}|D=1} = C_{\eta + V_{t-1}, \eta + V_{t-2}|D=1}$, the \ref{ass:copula-stability} holds.
\end{proposition}

The proof of \Cref{prop:twfe} is provided in Supplementary Appendix \ref{app:model-proofs}.  The additional condition in \Cref{prop:twfe} is fairly weak.  It allows for serial correlation in the time varying unobservables and for the distribution of time varying unobservables to change over time.  It also allows for the distribution of time varying unobservables to depend on the value of the individual heterogeneity.  There are also a number of special cases of this condition that are familiar as well.  One example is when $F_{V_{t},V_{t-1}|\eta,D=1} = F_{V_{t-1},V_{t-2}|\eta,D=1}$.  This says that the joint distribution of time varying unobservables does not change over time conditional on individual unobserved heterogeneity.  Another leading case is when $(V_t, V_{t-1}, V_{t-2}) \independent \eta | D=1$.  In this case, the \ref{ass:copula-stability} will hold as along as $C_{V_t,V_{t-1}|D=1} = C_{V_{t-1},V_{t-2}|D=1}$.  This latter condition will hold, for example, in the case where the $V_s$ are mutually independent.  These sorts of conditions are frequently invoked in the literature on treatment effects with panel data using identification arguments from the measurement error literature (e.g., \citet{li-vuong-1998,evdokimov-2010,bonhomme-sauder-2011,canay-2011,freyberger-2018}).

\subsubsection{Change in Changes}

In the application on job displacement, I use the Change in Changes (CIC) model (\citet{athey-imbens-2006}) to identify the counterfactual distribution of outcomes $F_{Y_{0t}|D=1}$ that individuals in the treated group would have experienced if they had not been displaced from their job.  Therefore, it is useful to consider what extra conditions need to be placed on the CIC model in order for the Copula Stability Assumption to hold.  The CIC model is based on the following setup:
\begin{align*}
  Y_{0is} = h_s(U_{is}) \quad \mathrm{for}\ s=t,t-1
\end{align*}
with (i) $h_s(u)$ strictly increasing in $u$ and (ii) $U_{s}|D=d \sim F_{U|D=d}$ for $s=t,t-1$ and $d=0,1$.  In addition to these conditions, I also assume $Y_{0it-2}=h_{t-2}(U_{it-2})$, $U_{is} = \eta_i + V_{is}$ for $s=t,t-1,t-2$, and $V_{s}|\eta,D=d \sim F_{V|\eta,D=d}$ for all $s=t,t-1,t-2$ (this last condition implies $U_{s}|D=d \sim F_{U|D=d}$ in each time period).  These extra conditions simply extend the model to three periods and from the case with repeated cross sections to panel data.  Importantly, they allow the distribution of $\eta$ to differ across the treated and untreated group.  I call the set of conditions above the Three Period Panel CIC model.    
\begin{proposition} \label{prop:2}  In the Three Period Panel CIC model given above, and under the additional assumption that $F_{V_t,V_{t-1}|\eta,D=1}=F_{V_{t-1},V_{t-2}|\eta,D=1}$, the \ref{ass:copula-stability} holds.
\end{proposition}

The proof of \Cref{prop:2} is provided in Supplementary Appendix \ref{app:model-proofs}.  Proposition \ref{prop:2} gives the additional condition required for the \ref{ass:copula-stability} to hold in the CIC model.  The additional condition is weak and would be satisfied if the $V_s$ are iid, but it can also allows for dependence of $V_s$ over time as long as the dependence is constant.

\begin{remark}
  Both the two way fixed effects model and the Change in Changes model allow for individuals to select into treatment (i) based on having ``good'' treated potential outcomes and (ii) based on their unobserved heterogeneity, $\eta_i$, that shows up in the model for untreated potential outcomes.\footnote{The first part holds because the models do not put any structure on how treated potential outcomes are generated, and the second part holds because the unobserved heterogeneity can be distributed differently for individuals in the treated group relative to individuals in the untreated group.}  The additional conditions for the Copula Stability Assumption to hold in each model allow for this type of selection into treatment as well.  
\end{remark}

\begin{remark} \label{rem:ri}
  Neither of the rank invariance assumptions mentioned in Section 2 are likely to hold in the two way fixed effects model nor in the Change in Changes model.  Cross-sectional rank invariance is related to the model for treated potential outcomes (which was not specified in either case above) as well as the model for untreated potential outcomes, but it would take an unusual model for treated potential outcomes to generate cross-sectional rank invariance.  Rank invariance over time is not compatible with the time varying shocks that show up in both models given in this section -- these cause individuals to change their ranks in the distribution of outcomes over time.
\end{remark}

\begin{remark}
  One case where the \ref{ass:copula-stability} does not hold, in general, is when untreated potential outcomes are generated by an interactive fixed effects model; e.g., $Y_{0is} = \theta_s + \eta_i + \lambda_i F_s + V_{is}$ where the notation is the same as for the two-way fixed effects model above and, in addition, $\lambda_i$ is unobserved heterogeneity whose effect, $F_s$, can change over time.  The reason that the \ref{ass:copula-stability} does not hold in this case is that, when $F_s$ can vary arbitrarily over time, the dependence between untreated potential outcomes over time can vary in essentially unrestricted ways over time.\footnote{Interestingly, however, if $F_s$ changes monotonically over time, the bounds coming from the \ref{ass:copula-stability} will be conservative (here, suppose for simplicity that the $V_{is}$ are iid).  This holds because, when $F_s$ is changing monotonically over time, there will be increasing dependence in untreated potential outcomes over time.  Then, the result in \Cref{prop:1} implies that the resulting bounds will be conservative.  Monotonicity of $F_s$ is reasonable in many applications where the researcher thinks that, for example, the effect of some unobserved variable is increasing or decreasing over time.  It will also hold in special cases of the interactive fixed effects model such as individual-specific linear trends models that are common in applied work as well.}
\end{remark}

\subsection{Pre-Testing the Copula Stability Assumption}

In treatment effects applications with more periods than are required for identification, empirical researchers frequently ``pre-test'' their identifying assumptions (see, for example, \citet{callaway-santanna-2019,roth-2018}).  The idea is that, even though the identifying assumptions themselves are not directly testable, it is often reasonable to think that the identifying assumptions would have held in earlier periods as well.  Then, one can compare results using observed untreated potential outcomes in pre-treatment periods to results coming from the identifying assumptions.  In the current case, when there are extra pre-treatment time periods, one can use the same idea to pre-test the \ref{ass:copula-stability} by testing if the copula of untreated potential outcomes over time does not change over time in pre-treatment periods.  In Supplementary Appendix \ref{app:pretest}, I propose a nonparametric version of this pre-test building on results from the goodness-of-fit testing literature.  Another simple alternative is to compute some dependence measure such as Spearman's Rho or Kendall's Tau and test if it remains constant across pre-treatment periods.  This is a practical alternative in the sense that it is easy to implement though it could fail to detect certain violations of the \ref{ass:copula-stability} in pre-treatment periods.

\section{Estimation and Inference}

This section proposes estimators for the DoTT and QoTT, provides the limiting distribution of these estimators, and shows the validity of the numerical bootstrap of \citet{hong-li-2018} to conduct inference.  This section explicitly conditions on covariates $X$ in order to increase the clarity of the arguments though note that all of the unconditional results hold simply by taking the covariates to only include a constant.  The proofs for all the results in this section are provided in Supplementary Appendix \ref{app:asymptotic}.  At a high level, the estimators of the DoTT and QoTT come from plugging in first step estimators into the expressions given in \Cref{theorem:dte,theorem:qotet}.  In particular,
\begin{align*}
  \widehat{DoTT}^L(\delta) = \frac{1}{n} \sum_{i=1}^n \frac{D_i}{p} \ \hat{F}^L_{Y_{1t}-Y_{0t}|Y_{0t-1},X,D=1}(\delta|Y_{it-1},X_i) \\[15pt]
  \widehat{DoTT}^U(\delta) = \frac{1}{n} \sum_{i=1}^n \frac{D_i}{p} \ \hat{F}^U_{Y_{1t}-Y_{0t}|Y_{0t-1},X,D=1}(\delta|Y_{it-1},X_i)
\end{align*}
where $p$ is the fraction of individuals in the treated group.\footnote{The term $D_i/p$ makes these terms equivalent to averaging over observations in the treated group only.}  The estimators of $F^L_{Y_{1t}-Y_{0t}|Y_{0t-1},X,D=1}$ and $F^U_{Y_{1t}-Y_{0t}|Y_{0t-1}, X,D=1}$ further depend on preliminary estimators of $F_{Y_{1t}|Y_{0t-1},X,D=1}$ and $F_{Y_{0t}|Y_{0t-1},X,D=1}$ (see \Cref{lemma:cond-dte}).  $F_{Y_{1t}|Y_{0t-1},X,D=1}$ is identified by the sampling process and can be estimated directly; on the other hand, $F_{Y_{0t}|Y_{0t-1},X,D=1}$ requires further preliminary estimators of $F_{Y_{0t}|X,D=1}$, $F_{Y_{0t-1}|X,D=1}$, and $F_{Y_{0t-2}|X,D=1}$ (see \Cref{lem:idy0}).  Recall that $F_{Y_{0t}|X,D=1}$, which is the counterfactual distribution of untreated potential outcomes for individuals in the treated group, requires an identifying assumption.  I use a conditional version of the Change in Changes model (\citet{melly-santangelo-2015}) to identify this distribution.\footnote{As mentioned in Footnote \ref{fn:idy0}, the assumptions used to identify the counterfactual distribution of untreated potential outcomes for individuals in the treated group may have identifying power for the joint distribution of treated and untreated potential outcomes.  For the Change in Changes model, this is not the case.  The reasons are that (i) this setup does not restrict how treated potential outcomes are generated at all, and (ii) it is consistent with any dependence structure between $Y_{0t}$ and $Y_{0t-1}$.  Other models for untreated potential outcomes with panel data (e.g., Quantile Difference in Differences or the method proposed in \citet{bonhomme-sauder-2011}) do not imply any restrictions on the joint distribution of potential outcomes either.  This is not always the case though.  For example, if one assumes selection on observables (i.e., that treatment is as good as randomly assigned once one conditions on a lag of the outcome), that would identify the joint distributions of $(Y_{1t},Y_{0-1})|D=1$ and $(Y_{0t},Y_{0t-1})|D=1$ which would lead to the same bounds as the ones in the current paper without requiring the \ref{ass:copula-stability}.  That being said, in economics, models like the Change in Changes model are more common because they can be motivated using models that allow for an important role to be played by unobserved heterogeneity.  For example, in the context of linear models, models such as Change in Changes are closely related to fixed effects models, while selection on observables is more similar to regressions that include lagged outcomes but not fixed effects. \label{fn:ass2} }   Under this setup, the counterfactual distribution is given by 
\begin{align*}
  F_{Y_{0t}|X,D=1}(y|x) = F_{Y_{0t-1}|X,D=1}(F^{-1}_{Y_{0t-1}|X,D=0}(F_{Y_{0t}|X,D=0}(y|x)|x)|x)
\end{align*}
Thus, to estimate the bounds on the DoTT and QoTT requires preliminary estimators of (i) $F_{Y_{1t}|Y_{0t-1},X,D=1}$ and $F_{Y_{0t}|Y_{0t-1},X,D=1}$ as well as (ii) $F_{Y_{0t-1}|X,D=1}$, $F_{Y_{0t-2}|X,D=1}$, $F_{Y_{0t}|X,D=0}$, and $F_{Y_{0t-1}|X,D=0}$.  I propose using flexible parametric first step estimators based on distribution regression (for the first group) and quantile regression (for the second group).\footnote{ For these, I proceed by estimating all of their conditional quantiles using quantile regression and then inverting to obtain the conditional distribution.}  Similar approaches have been used in \citet{chernozhukov-val-melly-2013,melly-santangelo-2015,wuthrich-2019}.\footnote{Using flexible parametric first step estimators for conditional quantiles and distributions is an attractive option in many applications.  They are substantially more flexible than imposing that outcomes follow a particular distribution that is known up to a few parameters (e.g., normal), but they are also more feasible than nonparametric estimators that require extra regularity conditions and may be difficult to implement in many applications in economics.  This is especially true in the frequently encountered case with a relatively large number of covariates and only a moderate number of observations.}

Next, I develop some asymptotic theory for the estimators of the DoTT and QoTT.    %
At a high level, the arguments in this section proceed by first showing that distribution regression estimators $F_{Y_{1t}|Y_{0t-1},X,D=1}$ and $F_{Y_{0t}|Y_{0t-1},X,D=1}$ converge uniformly to Gaussian processes.  This step requires proving some new results on distribution regression estimators when one of the regressors is ``generated'' as well as when the index depends on an estimated transformation (see \Cref{app:dr}).  Second, building on work on inference on parameters that depend on the joint distribution of potential outcomes (\citet{fan-park-2010,fan-wu-2010}) as well as recent results in the literature on Hadamard directionally differentiable functions (e.g, \citet{fang-santos-2019}), one can derive the limiting distribution of the upper and lower bound of the DoTT and QoTT.  To conduct inference, in practice, the empirical bootstrap can be applied to simulate the limiting distribution of the first step estimators which can be combined with the numerical delta method (\citet{hong-li-2018}) to simulate the limiting distributions of estimators of the bounds.\footnote{It is also worth mentioning that I focus on constructing pointwise confidence intervals for the bounds on the $DoTT$ and $QoTT$.  One main reason to consider pointwise confidence intervals is that the bounds themselves are pointwise (see the discussion in \citet{firpo-ridder-2019}).  However, it does seem possible to extend these results to uniform confidence bands using, for example, the approach in \citet{masten-poirier-2020} and exploiting the monotonicity of the bounds.  I also focus on inference for the bounds themselves though one could also consider inference on the $DoTT$ or $QoTT$ itself rather than the bounds (see, for example, \citet{fan-park-2012,chernozhukov-lee-rosen-2013}).}

For $(d,s) \in \{0,1\} \times \{t,t-1,t-2\}$, let 
$\hat{G}_{d,s}(y,x) = \sqrt{n}(\hat{F}_{Y_s|X,D=d}(y|x) - F_{Y_s|X,D=d}(y|x))$, and let $\hat{G}^0(y,x) = \sqrt{n}(\hat{F}_{Y_{0t}|X,D=1}(y|x) - F_{Y_{0t}|X,D=1}(y|x))$.
Let $\mathcal{Y}_{ds}$, $\mathcal{X}_d$, and $\mathbf{\Delta}$ denote the supports of $Y$ for group $d$ in time period $s$, $X$ for group $d$, and $(Y_{1t}-Y_{0t})|D=1$, respectively.  Also, let $\bar{\mathcal{Y}}_{0t}$ denote the support of $Y_{0t}|D=1$.  Also, for some set $S$, let $l^\infty(S)$ denote the space of all uniformly bounded functions on $S$ that are equipped with the supremum norm and $\mathcal{C}(S)$ denote the space of all uniformly continuous functions on $S$.  As a first step, I assume that a functional central limit theorem holds jointly for each of the first step estimators, and that they each converge uniformly at the parametric rate.

\begin{assumption}[Functional Central Limit Theorem for First-Step Estimators]\label{ass:fclt}
  \begin{align*}
    \sqrt{n}(\hat{G}_{1,t}, \hat{G}_{1,t-1}, \hat{G}_{1,t-2}, \hat{G}_{0,t}, \hat{G}_{0,t-1},\hat{G}^0) \rightsquigarrow (\mathbb{W}_{1,t}, \mathbb{W}_{1,t-1}, \mathbb{W}_{1,t-2}, \mathbb{W}_{0,t}, \mathbb{W}_{0,t-1}, \mathbb{W}^0)
  \end{align*}
  in the space $\mathcal{S} = l^\infty(\mathcal{Y}_{1t}\mathcal{X}_1) \times l^\infty(\mathcal{Y}_{1t-1} \mathcal{X}_1) \times l^\infty(\mathcal{Y}_{1t-2} \mathcal{X}_1) \times l^\infty(\mathcal{Y}_{0t}, \mathcal{X}_0) \times l^\infty(\mathcal{Y}_{0t-1}\mathcal{X}_0) \times l^\infty(\bar{\mathcal{Y}}_{0t}\mathcal{X}_1)$ where $(\mathbb{W}_{1,t}, \mathbb{W}_{1,t-1}, \mathbb{W}_{1,t-2}, \mathbb{W}_{0,t}, \mathbb{W}_{0,t-1}, \mathbb{W}^0)$ is a tight, mean zero Gaussian process.
\end{assumption}

\Cref{ass:fclt} says that first step estimators of the distribution of the outcomes in periods $t$, $t-1$, and $t-2$ and for both the treated and untreated groups as well as the counterfactual distribution of untreated potential outcomes for the treated group converge uniformly to a Gaussian process at the parametric rate.  In the application, I estimate the distributions of outcomes conditional on covariates for each time period and each group using quantile regression.  Therefore, the part of \Cref{ass:fclt} that involves observed outcomes holds under standard regularity conditions on quantile regression estimators (see \Cref{ass:qr} in Supplementary Appendix \ref{app:supp} for these additional conditions).  %
  The reason to state \Cref{ass:fclt} as an assumption rather than as a result is that it imposes a generic limiting process for the estimator of $F_{Y_{0t}|X,D=1}$.  The limiting process will depend on the particular identifying assumptions that are invoked for this distribution.  In Supplementary Appendix \ref{app:supp}, I discuss the particular case where this distribution is identified using Change in Changes as in \citet{athey-imbens-2006,melly-santangelo-2015}, but, as noted above, other approaches could be used as well.

A key intermediate step is to establish the limiting distributions of estimators of $F_{Y_{1t}|Y_{0t-1},X,D=1}$ and $F_{Y_{0t}|Y_{0t-1},X,D=1}$ (see \Cref{theorem:dte}).  I propose estimating each of these using distribution regression (\citet{foresi-peracchi-1995,chernozhukov-val-melly-2013}).  Handling $F_{Y_{1t}|Y_{0t-1},X,D=1}$ is straightforward as it amounts to a standard distribution regression of $Y_{1t}$ on $Y_{0t-1}$ and $X$ which are all observed for individuals in the treated group, and the results on limiting processes for distribution regression estimators from \citet{chernozhukov-val-melly-2013} can be applied directly.  In particular, it immediately follows from these results that 
\begin{align*}
  \sqrt{n} (\hat{F}_{Y_{1t}|Y_{0t-1},X,D=1} - F_{Y_{1t}|Y_{0t-1},X,D=1}) \rightsquigarrow \mathbb{G}_1
\end{align*}
where the exact conditions, which are standard, are given in Supplementary Appendix \ref{app:supp}, and the exact expression for $\mathbb{G}_1$ is given in \Cref{eqn:G1} in Supplementary Appendix \ref{app:supp}.  However, $F_{Y_{0t}|Y_{0t-1},X,D=1}$ is more challenging as it depends on applying estimated transformations to $Y_{0t-1}$ and $Y_{0t-2}$ before carrying out the distribution regression.  Next, I establish the limiting process for a distribution regression estimator for this case.

\begin{theorem}  \label{thm:fclt2} Under the \ref{ass:copula-stability}, \Cref{ass:id1,ass:id2,ass:id3,ass:fclt,ass:compact,ass:cont,ass:dr},
  \begin{align*}
    \sqrt{n}\begin{pmatrix}\hat{F}_{Y_{1t}|Y_{0t-1},X,D=1} - F_{Y_{1t}|Y_{0t-1},X,D=1} \\  \hat{F}_{Y_{0t}|Y_{0t-1},X,D=1} - F_{Y_{0t}|Y_{0t-1},X,D=1} \\ \hat{F}_{Y_{0t-1},X|D=1} - F_{Y_{0t-1},X|D=1} \end{pmatrix} \rightsquigarrow \begin{pmatrix} \mathbb{G}_1 \\ \mathbb{G}_2 \\ \mathbb{G}_3 \end{pmatrix}
  \end{align*}
  in the space $l^\infty(\mathcal{Y}_{1t}\mathcal{Y}_{1t-1}\mathcal{X}_1) \times l^\infty(\bar{\mathcal{Y}}_{0t} \mathcal{Y}_{1t-1} \mathcal{X}_1) \times l^\infty(\mathcal{Y}_{1t-1} \mathcal{X}_1)$ where $(\mathbb{G}_1, \mathbb{G}_2, \mathbb{G}_3)$ is a tight, mean zero Gaussian process and where $\mathbb{G}_2$ is defined in \Cref{eqn:G2} in \Cref{app:asymptotic}, and $\mathbb{G}_3$ is defined in \Cref{eqn:G3} in \Cref{app:supp}.
\end{theorem}

\Cref{thm:fclt2} provides the limiting process for the second step estimators.  These are important inputs into estimating the DoTT and QoTT.  In particular, the lower and upper bounds on the $DoTT$ and $QoTT$ can be viewed as functionals of the distributions of $F_{Y_{1t}|Y_{0t-1},X,D=1}$, $F_{Y_{0t}|Y_{0t-1},X,D=1}$ and $F_{Y_{0t-1},X|D=1}$; see \Cref{theorem:dte,theorem:qotet}.  One can define the bounds of the DoTT and QoTT as maps from these distributions.  Then, the key step for the main asymptotic results in this section is to show that this map is Hadamard directionally differentiable.

The final results of this section provide the limiting distribution for the estimators of the lower and upper bounds on the $DoTT$ and $QoTT$ in the main text.

\begin{proposition} \label{prop:cltdott}Under the \ref{ass:copula-stability}, \Cref{ass:id1,ass:id2,ass:id3,ass:fclt,ass:compact,ass:cont,ass:dr},
\begin{align*}
  \sqrt{n}(\widehat{DoTT}^L(\delta) - DoTT^L(\delta)) \rightsquigarrow \mathbb{V}^L \quad \textrm{and} \quad   \sqrt{n}(\widehat{DoTT}^U(\delta) - DoTT^U(\delta)) \rightsquigarrow \mathbb{V}^U 
\end{align*}
where explicit expressions for $\mathbb{V}^L$ and $\mathbb{V}^U$ are provided in Supplementary Appendix \ref{app:asymptotic}.
\end{proposition}

Finally, the main asymptotic results for the estimators of the bounds on the $QoTT$ follow under a small extension to the results in \Cref{prop:cltdott}, and I state these results next.

\begin{theorem} \label{thm:cltqott}
  Under the \ref{ass:copula-stability}, \Cref{ass:id1,ass:id2,ass:id3,ass:fclt,ass:compact,ass:cont,ass:dr} and for some $\tau$ satisfying $0 < \epsilon < \tau < (1-\epsilon) < 1$ for some $\epsilon>0$,
  \begin{align*}
    \sqrt{n}\left(\widehat{QoTT}^L(\tau) - QoTT^L(\tau)\right) \rightsquigarrow \mathbb{Z}^L \quad \textrm{and} \quad \sqrt{n}\left(\widehat{QoTT}^U(\tau) - QoTT^U(\tau)\right) \rightsquigarrow \mathbb{Z}^U
  \end{align*}
  where 
  \begin{align*}
    \mathbb{Z}^L = \frac{\mathbb{V}^U(QoTT^L(\tau))}{f_{DoTT^U}(QoTT^L(\tau))} \quad \textrm{and} \quad 
    \mathbb{Z}^U = \frac{\mathbb{V}^L(QoTT^U(\tau))}{f_{DoTT^L}(QoTT^U(\tau))}
  \end{align*}
  where $f_{DoTT^U}$ and $f_{DoTT^L}$ denote the densities of the upper and lower bounds on the $DoTT$, respectively.

\end{theorem}

\bigskip

The above results derive the limiting process for estimators of the bounds of the DoTT and QoTT.  As an intermediate result (see the proof of \Cref{prop:cltdott} in \Cref{app:asymptotic}), I show that the DoTT and QoTT are Hadamard directionally differentiable functionals of the the conditional distributions discussed in \Cref{thm:fclt2}.  However, they are not fully Hadamard differentiable.  The results in \citet{fang-santos-2019} thus imply that the standard empirical bootstrap cannot be used to conduct inference.  Instead, to conduct inference in practice, I use the numerical bootstrap method proposed in \citet{hong-li-2018}.  In particular, one can proceed by constructing bootstrap estimates of the first step estimators (here, using the standard empirical bootstrap), denote these by $\hat{F}^* := (\hat{F}^*_{Y_{1t}|Y_{0t-1},X,D=1}, \hat{F}^*_{Y_{0t}|Y_{0t-1},X,D=1}, \hat{F}^*_{Y_{0t-1},X|D=1})$ and denote the first step estimators themselves by $\hat{F} := (\hat{F}_{Y_{1t}|Y_{0t-1},X,D=1}, \hat{F}_{Y_{0t}|Y_{0t-1},X,D=1}, \hat{F}_{Y_{0t-1},X|D=1})$.  Here, generically, let $\phi$ denote one of the maps from distribution functions to a bounds parameter of interest, and let $\mathbb{V}$ generically represent the corresponding limiting distribution in either \Cref{prop:cltdott} (for the DoTT) or \Cref{thm:cltqott} (for the QoTT).  Let $\epsilon_n$ denote a tuning parameter that satisfies the conditions that $\epsilon_n \rightarrow 0$ and $\epsilon_n \sqrt{n} \rightarrow \infty$ as $n \rightarrow \infty$.  The results in \citet{hong-li-2018} imply that
\begin{align} \label{eqn:bstrap}
  \frac{\phi\big(\hat{F} + \epsilon_n \sqrt{n}(\hat{F}^* - \hat{F})\big) - \phi(\hat{F})}{\epsilon_n} \rightsquigarrow \mathbb{V}
\end{align}
Essentially, the idea here is to combine a numerical derivative of the map $\phi$ with bootstrapping the first step estimators.\footnote{It is also helpful to understand the term on the left hand side of \Cref{eqn:bstrap} by noticing that it corresponds to the standard empirical bootstrap if $\epsilon_n = n^{-1/2}$ (though this case is ruled out by the rate conditions placed on $\epsilon_n$).} The term on the left hand side of \Cref{eqn:bstrap} can be simulated a large number of times, and the resulting distribution will approximate the limiting distribution of whichever bounds parameter one is interested in.  In the application, I report confidence intervals for the bounds on DoTT and QoTT.  For a lower $(1-\alpha)$ confidence interval on $DoTT^L(\delta)$ or $QoTT^L(\tau)$, one can compute $\phi^L(\hat{F}) - \hat{c}^L_{1-\alpha}/\sqrt{n}$ where $\hat{c}^L_{1-\alpha}$ is the $(1-\alpha)$-quantile of the simulated version of the term on the left hand side of \Cref{eqn:bstrap} and $\phi^L$ generically denotes the map from distribution functions to either $DoTT^L$ or $QoTT^L$.  Similarly, an upper $(1-\alpha)$ confidence interval for $DoTT^U(\delta)$ or $QoTT^U(\tau)$ is given by $\phi^U(\hat{F}) - \hat{c}^U_{\alpha}/\sqrt{n}$ where $\hat{c}^U_{\alpha}$ is the $\alpha$-quantile of the simulated version of the term on the left hand side of \Cref{eqn:bstrap} and $\phi^U$ denotes the map from distribution functions to either $DoTT^U(\delta)$ or $QoTT^U(\tau)$.  This is the approach that I take to conducting inference in the application.

\section{Job Displacement during the Great Recession}

This section studies the effect of job displacement during the Great Recession on yearly earnings of late prime-age workers.  The main goal here is to try to learn about features of the distribution of the individual-level effect of job displacement.  That is, unlike the vast majority of work on job displacement which estimates an average effect of job displacement,\footnote{Supplementary Appendix \ref{app:jd-details} discusses existing work on job displacement in more detail.} this part of the paper tries to learn about how the effect of job displacement varies across different individuals.  Like much of the literature on job displacement, I find a large negative average effect of job displacement.  In addition, though, the evidence in this section indicates that there is a large amount of heterogeneity across displaced workers where the average effect combines very large negative effects of job displacement for some workers with much more modest effects of job displacement for other workers.%

\subsection{Data and Baseline Results}

The data comes from the 1979 National Longitudinal Survey of Youth (NLSY).  The main outcome variable is the log of yearly earnings in 2011.  In 2011, NLSY respondents are between 47 and 54 years old.  I limit the sample to individuals who worked at least 1000 hours in 2007 and classify workers as being displaced if they left a job in 2008 or 2009 and the reason given is (i) layoff, job eliminated or (ii) company, office, or workplace closed.  This excludes other reasons for leaving a job such as being fired, quitting, moving, or the end of temporary employment.\footnote{A main issue in the job displacement literature is how to treat individuals with zero earnings.  Most research drops individuals with zero earnings (for example, research that uses state-level administrative data cannot tell the difference between actually having zero earnings and moving to another state).  Even most research on job displacement using the PSID or NLSY tends to drop individuals with zero earnings, then take the log of earnings as the outcome variable in some regression (\citet{stevens-1997, kletzer-fairlie-2003}).  I follow this same approach and restrict the sample to individuals who have at least \$1,000 in earnings in 2003, 2007, and 2011.}  The sample includes 2,775 individuals, of whom 122 are displaced in 2008 or 2009.

Summary statistics are provided in Table \ref{tab:ss}.  Displaced workers and non-displaced workers have similar education levels.  There are no statistically significant differences in the fraction that have less than high school education, graduated from high school, or graduated from college.  
On the other hand, there are larger differences in race; displaced workers are 10 percentage points more likely to be black (p-value: 0.02).  

Table \ref{tab:ss} also contains the path of average earnings for displaced and non-displaced workers from 2001-2013.  In 2001, average yearly earnings for displaced workers was about \$43,000 while average earnings for non-displaced workers was about \$47,000 (p-value of the difference: 0.40).  From 2001-2007, which are all pre-displacement years, the gap remains roughly constant and does not display any obvious trend over time.  However, by 2009, displaced workers have experienced a large decline in average earnings.  The gap in 2009 is roughly \$9,000 and almost all of this is explained by the decline in earnings of displaced workers.  By 2011, the gap in earnings has increased to almost \$22,000; about a third of the increase in the gap is due to increased earnings of non-displaced workers and the remaining part of the increase in the gap is due to decreased earnings of displaced workers.\footnote{Recall that workers are displaced in either 2008 or 2009, so earnings in 2009 may mix pre-displacement earnings for workers who are displaced at some point in 2009 with their post-displacement earnings which may account for the smaller earnings gap in 2009 than in 2011 or 2013.}  The gap is still close to \$20,000 in 2013.

Next, I estimate the counterfactual distribution of untreated potential outcomes for the treated group -- i.e., the unobserved distribution of earnings for the group of displaced workers if they had not been displaced -- using the Change in Changes method of \citet{athey-imbens-2006}.  Knowledge of this distribution, in combination with the distribution of treated potential outcomes for the treated group (which is observed), identifies the ATT and the QTT.\footnote{\label{fn:pc}In this part of the paper, the outcome is the logarithm of earnings.  However, because many of the effects are relatively large, I convert estimated effects from ``log points'' into percentage changes using $\exp(\hat{\alpha}) - 1$ where $\hat{\alpha}$ is some estimated parameter of interest in log points.}  The results indicate that late prime-age workers lose 34\% of their earnings on average due to job displacement.\footnote{This effect is quite similar in magnitude compared to estimates of the effect on prime age workers during the deep recession in the early 1980s (\citet{jacobson-lalonde-sullivan-1993, wachter-song-manchester-2009}) which are the largest in the literature.  This estimate is broadly similar to the estimated effect of job displacement on all workers during the Great Recession reported in \citet{farber-2017} and using the Displaced Workers Survey.}%

Estimates of both marginal distributions are presented in \autoref{fig:marg-dists}, and an estimate of the QTT is presented in \autoref{fig:qtet} (both of these figures are available in the Supplementary Appendix).\footnote{There are a variety of other methods that could be used for the first step estimation instead of Change in Changes including the Panel DID method of \textcite{callaway-li-2019}, Quantile Difference in Differences (\citet{athey-imbens-2006}), or by assuming selection on observables holds after conditioning on a lag of earnings (\citet{firpo-2007}).  Figure \ref{fig:qte-plots} (in the Supplementary Appendix) plots the QTT under each of these alternative assumptions.  The QTT results are not sensitive to changes in the underlying model for the counterfactual distribution of untreated potential outcomes.  This result also implies that the results for parameters that depend on the joint distribution of potential outcomes are not sensitive to the choice of identifying assumption in the first stage.}  The QTT is negative everywhere and statistically significant for all quantiles except the very largest (primarily due to standard errors increasing).  The differences between the marginal distributions of observed outcomes for displaced workers and the distribution of outcomes that displaced workers would have experienced if they had not been displaced are large.  The QTT is also increasing across quantiles.  It is tempting to interpret Figure \ref{fig:qtet} as indicating that the effect of job displacement is largest for individuals at the lower part of the earnings distribution.  However, as mentioned above, job displacement appears to cause many individuals to change their rank in the distribution of earnings which makes this interpretation of the QTT invalid.  %

\subsection{Distributional Effects of Job Displacement}

Next, I use the techniques presented earlier in the paper to estimate some parameters that depend on the joint distribution of treated and untreated potential outcomes.  First, I consider the QoTT.  \Cref{fig:wcb} plots bounds on the QoTT under no assumptions on the dependence between potential outcome distributions.  
There are several things to notice from the figure.  First, immediately these bounds imply that there are heterogeneous effects -- the upper bound on the 5th percentile of the effect of job displacement is estimated to be 56\% lower earnings while the lower bound on the 95th percentile of the effect of job displacement is estimated to be 20\% lower earnings.  Finding that there is treatment effect heterogeneity is important here as it implies that the effect of job displacement can potentially be more severe for some displaced workers than the average effect would indicate.  These bounds also imply that at least 5.9\% of displaced workers (and up to 63.0\% of displaced workers) lose more than half of their earnings relative to what they would have earned if they had not been displaced.   However, these bounds are relatively less informative about other parameters of interest.  For example, the median of the treatment effect is bounded to be between an earnings loss of 61\% and an earnings gain of 65\%.  
The bounds are consistent with up to 81\% of displaced workers having higher earnings following displacement than they would have had if they had not been displaced, but they are also consistent with all displaced workers having lower earnings following displacement than they would have had if they had not been displaced.  To better understand heterogeneous effects of job displacement, it would be helpful to have tighter bounds on these parameters.  
It is also worth mentioning that, although the confidence intervals appear small in the \Cref{fig:wcb}, it is actually more the case that the width of the bounds stemming from the identification arguments is wide relative to the uncertainty due to estimation.  For example, as mentioned above, the point estimate for the upper bound of the median of the treatment effect is 65\% higher earnings, and the corresponding one-sided upper 95\% confidence interval is 79\% higher earnings.

Next, \Cref{fig:csab} provides bounds under the \ref{ass:copula-stability}. These bounds are indeed tighter.  Earnings losses at the 5th percentile are between 73\% and 96\% which implies that, at least for some individuals, the effect of job displacement is much worse than the average effect.  Similarly, the bounds imply that at least 9.0\% of displaced workers earn less than half of what they would have earned if they had not been displaced.  Interestingly, the bounds also imply that at least 10.1\% of individuals have higher earnings after being displaced than they would have had they not been displaced (the corresponding one-sided lower confidence interval is 4.0\%).  This type of conclusion was not available without exploiting the \ref{ass:copula-stability} and implies that the \ref{ass:copula-stability} is incompatible with the assumption of Monotone Treatment Response in the current case.  This result provides another indication that the effect of job displacement is quite heterogeneous across workers.

Next, it is interesting to compare the results in the application to results that exploit having access to covariates.  As discussed above, covariates can be used to tighten bounds on distributional treatment effect parameters, and they can also be used to tighten the bounds under the \ref{ass:copula-stability}.  Here, I use education levels, race, and gender as covariates.  These results are available in \Cref{fig:compare-bounds}.  Covariates do indeed tighten the bounds relative to the bounds that only use information on the marginal distributions of treated and untreated potential outcomes (this can be seen in the bottom-right panel of \Cref{fig:compare-bounds}); however, overall the effect is fairly moderate -- the upper bound on the QoTT becomes somewhat tighter and the lower bound is basically the same.  When covariates are used to further tighten the bounds that come from the \ref{ass:copula-stability}, essentially the same pattern holds -- the upper bound on the QoTT becomes somewhat tighter and the lower bound remains essentially the same.  Finally, it is useful to compare the bounds (i) under the \ref{ass:copula-stability} without covariates to the bounds using (ii) covariates but not the \ref{ass:copula-stability}.  Here, the bounds that use the \ref{ass:copula-stability} are noticeably tighter.  This indicates that, at least in the current application, exploiting panel data through the \ref{ass:copula-stability} is particularly useful in deriving tighter bounds on the distributional treatment effects that depend on the joint distribution of potential outcomes.

Finally, in Supplementary Appendix \ref{app:jd-details}, I compare the bounds under the \ref{ass:copula-stability} to results under the assumptions of cross-sectional rank invariance and rank invariance over time.  In the current application, the \ref{ass:copula-stability} is incompatible with the assumption of cross-sectional rank invariance (see \Cref{fig:qott-ri-nlsy}); in particular, the bounds under the \ref{ass:copula-stability} imply that there is more treatment effect heterogeneity than is delivered under the assumption of cross-sectional rank invariance.  On the other hand, the point estimates for the QoTT under rank invariance over time fall completely within the bounds the \ref{ass:copula-stability}.  Despite this, the assumption of rank invariance over time can be pre-tested, and it is rejected in all pre-treatment time periods.  See Supplementary Appendix \ref{app:jd-details} for a more detailed discussion.

\paragraph{Does the \ref{ass:copula-stability} hold for job displacement?} \mbox{}

As a final step in the analysis, I pre-test the \ref{ass:copula-stability}.  First, Figure \ref{fig:spearmans-rho} plots Spearman's Rho for $Y_{0s}$ and $Y_{0s-1}$ in periods before the Great Recession separately for the treated group and untreated group (if the \ref{ass:copula-stability} holds in pre-treatment periods, then Spearman's Rho should be constant over time).  Here, as above, periods are earnings separated by 4 years starting from 1987 and going to 2011.\footnote{For the results in this section, I refer to the copula (or Spearman's Rho) between earnings in period $s-1$ and $s$ as the copula (or Spearman's Rho) in period $s$.  For example, the phrase ``Spearman's Rho in 1991'' refers to Spearman's Rho of earnings in 1987 and 1991.}  Spearman's Rho is broadly similar for both groups.  For the group of displaced workers, Spearman's Rho is increasing from 1991 (about 0.5) to 1999 (about 0.75) and is essentially constant from 1999 - 2007.  In 1987, individuals in the NLSY were between 23 and 30 years old.  In 1999, they were between 35 and 42 years old.  Figure \ref{fig:spearmans-rho} suggests that individuals moved around in the distribution of earnings more in their 20s, that there is less earnings mobility starting in their 30s and that the amount of mobility is the same through their 40s and early 50s.  That Spearman's Rho is constant from 1999-2007 provides a strong piece of evidence in favor of the \ref{ass:copula-stability}.

In terms of a formal statistical test, I (marginally) fail to reject that Spearman's Rho is constant for the entire period (p-value: .09); using the nonparametric version of the pre-test for the \ref{ass:copula-stability} (see Supplementary Appendix \ref{app:pretest} for additional details), I also fail to reject that the copula does not change over time for displaced workers (p-value: 0.53).  When I conduct the same tests excluding the copula in 1991 (when displaced workers were still quite young), I fail to reject that the copula is the same over time using either test.\footnote{One potential explanation for these results is that the test lacks power due to a relatively small number of displaced workers.  Therefore, I tried the same test with the group of non-displaced workers.  The bounds in the current paper do not require the copula of outcomes over time for individuals in the untreated group to be stable, but like the pre-test, if this copula does not change over time, it does provide suggestive evidence in favor of the \ref{ass:copula-stability}.  The nonparametric test for the group of non-displaced workers rejects that the copula is the same across all available time periods (p-value: 0.00); on the other hand, the test fails to reject that the copula is the same over the three most recent time periods (p-value: 0.74).  The same results hold for a test based on Spearman's Rho as well; together, these results are essentially in line with the results presented in \Cref{fig:spearmans-rho}.}

\section{Conclusion}

There is a long history in economics of exploiting access to panel data to identify parameters of interest when there is unobserved heterogeneity that affects outcomes.  This paper has developed a new approach to deriving tighter bounds on distributional treatment effect parameters that depend on the joint distribution of potential outcomes in the presence of panel data.  The results depend on three key ingredients: (i) access to at least three periods of panel data, (ii) identification of the marginal distribution of untreated potential outcomes for the treated group and (iii) the \ref{ass:copula-stability} which says that the dependence between untreated potential outcomes over time does not change over time.  The last of these is the key idea that allows the researcher to exploit having access to panel data to learn about the joint distribution of potential outcomes.  This type of idea may also be useful in other cases where the researcher has access to panel data.

Using these methods, I have studied the distributional effects of job displacement during the Great Recession for late prime-age workers.  Using standard techniques, I find that these workers lose on average 34\% of their yearly earnings following job displacement.  Using the techniques developed in the current paper, I find that this average effect masks substantial heterogeneity:  some workers lose a very large fraction of their earnings following job displacement though at least some workers have higher earnings following displacement than they would have had if they had not been displaced.  Having access to panel data and using the approach in the paper led to substantially tighter bounds on the partially identified parameters considered in the paper.

\pagebreak

\singlespace
\printbibliography

\pagebreak

\appendix

\numberwithin{equation}{section}
\numberwithin{assumption}{section}
\numberwithin{theorem}{section}
\numberwithin{proposition}{section}
\numberwithin{lemma}{section}
\numberwithin{remark}{remark}

\section{Proofs}

\paragraph{Proof of \autoref{lemma:joint-cond}} \mbox{}

\autoref{lemma:joint-cond} follows by an application of the Fr\'{e}chet-Hoeffding bounds to a conditional bivariate distribution.

\paragraph{Proof of \autoref{lemma:cond-dte}} \mbox{}

\autoref{lemma:cond-dte} applies the sharp bounds on the difference between random variables with known marginal distributions but unknown copula of \textcite{williamson-downs-1990} (See also: \textcite{makarov-1982, ruschendorf-1982, frank-nelsen-schweizer-1987, fan-park-2010}) to the difference conditional on the previous outcome.

\paragraph{Proofs of  \autoref{theorem:joint} and \autoref{theorem:dte}} \mbox{}

\autoref{theorem:joint} and \autoref{theorem:dte} follow from results in \textcite[][Section 5]{fan-park-2010}, \textcite{fan-guerre-zhu-2017}, and \citet{firpo-ridder-2019} which derive bounds on the unconditional distribution of the treatment effect when conditional marginal distributions are known.  In those cases, the marginal distributions are conditional on observed covariates $X$; in the current paper, the marginal distributions are conditional on a lag of the outcome $Y_{0t-1}$.

\paragraph{Proof of \autoref{theorem:qotet}} \mbox{}

\autoref{theorem:qotet} holds because inverting sharp bounds on a distribution implies sharp bounds on the quantiles (\citet{williamson-downs-1990, fan-park-2010}).

\paragraph{Proof of Proposition \ref{prop:1}} \mbox{}

  To simplify notation throughout the proof, let $F_{1|0} = F_{Y_{1t}|Y_{0t-1}, D=1}$ and let $F_{t-1}= F_{Y_{0t-1}|D=1}$.

  \bigskip
  
  \subproof{Upper Bound:} \mbox{}

  Let $F_1^U(y_0,y_1)$ denote the upper bound on the joint distribution given in Theorem \ref{theorem:joint} when $F_{Y_{0t}|Y_{0t-1},D=1} = F_1$ and likewise let $F_2^U(y_0,y_1)$ denote the upper bound when $F_{Y_{0t}|Y_{0t-1},D=1} = F_2$ and under the conditions stated in the proposition.  The goal is to show that $F_1^U(y_0, y_1) - F_2^U(y_0, y_1) \geq 0$ for all $(y_0, y_1) \in \mathcal{Y} \times \mathcal{Y}$ where, for simplicity, I suppose that $Y_{1t}$, $Y_{0t}$, and $Y_{0t-1}$ have common support $\mathcal{Y}$ conditional on $D=1$.  I also suppose (as elsewhere in the paper) that $\mathcal{Y}$ is compact, and that $Y_{1t}$ and $Y_{0t}$ are uniformly continuously distributed conditional on $Y_{0t-1}$ and $D=1$.

  First, it is straightforward to show that
  \begin{align}
    & F_1^U(y_0, y_1) - F_2^U(y_0, y_1) \nonumber \\
    & \hspace{50pt} = \int_{\mathcal{Y}} (F_1(y_0|y') - F_2(y_0|y')) \indicator{ F_1(y_0|y') \leq F_{1|0}(y_1|y'), F_2(y_0|y') \leq F_{1|0}(y_1|y') } \d{F_{t-1}(y')} \nonumber\\
    & \hspace{50pt} + \int_{\mathcal{Y}} (F_1(y_0|y') - F_{1|0}(y_1|y')) \indicator{ F_1(y_0|y') \leq F_{1|0}(y_1|y'), F_2(y_0|y') > F_{1|0}(y_1|y') } \d{F_{t-1}(y')} \nonumber \\
    & \hspace{50pt} + \int_{\mathcal{Y}} (F_{1|0}(y_1|y') - F_2(y_0|y')) \indicator{ F_1(y_0|y') > F_{1|0}(y_1|y'), F_2(y_0|y') \leq F_{1|0}(y_1|y') } \d{F_{t-1}(y')} \label{eqn:prop1a}
  \end{align}
  which holds from taking differences in the the conditional bounds on the joint distribution as in \Cref{lem:idy0}.  
  Next, $F_{1|0} \prec^{SI} F_1 \prec^{SI} F_2$ implies that there exist $y_1^* = y_1^*(y_0, y_1)$ and $y_2^* = y_2^*(y_0,y_1)$ such that
  \begin{align*}
    F_{1|0}(y_1|y_1^*) = F_1(y_0|y_1^*) \quad \mathrm{and}\quad F_{1|0}(y_1|y_2^*) = F_2(y_0|y_2^*)
  \end{align*}
  and for $y' \in \mathcal{Y}$ such that 
  \begin{align*}
    \ y' \geq y_1^* \implies F_{1|0}(y_1|y') \geq F_1(y_0|y'), &\qquad \  y' < y_1^* \implies F_{1|0}(y_1|y') < F_1(y_0|y')
  \end{align*}
  and
  \begin{align*}
    \ y' \geq y_2^* \implies F_{1|0}(y_1|y') \geq F_2(y_0|y'), &\qquad \  y' < y_2^* \implies F_{1|0}(y_1|y') < F_2(y_0|y')
  \end{align*}
  Note that the set of $y' \in \mathcal{Y}$ satisfying $y' < y_j^*$, for $j=1,2$ can be empty (particularly in the case where $y_j^* = y_{min}$ where $y_{min}$ is the smallest value in $\mathcal{Y}$).  Plugging these into Equation \ref{eqn:prop1a} implies that the difference in Equation \ref{eqn:prop1a} is given by
  \begin{align}
 & \int_{\mathcal{Y}} (F_1(y_0|y') - F_2(y_0|y')) \indicator{ y' \geq y_1^*, y' \geq y_2^* } \d{F_{t-1}(y')} \label{eqn:a} \\
    & \hspace{50pt} + \int_{\mathcal{Y}} (F_1(y_0|y') - F_{1|0}(y_1|y')) \indicator{ y_1^* \leq y' < y_2^* } \d{F_{t-1}(y')}  \label{eqn:b} \\
    & \hspace{50pt} + \int_{\mathcal{Y}} (F_{1|0}(y_1|y') - F_2(y_0|y')) \indicator{ y_2^* \leq y' < y_1^*} \d{F_{t-1}(y')} \label{eqn:c}
  \end{align}

  \textit{Case 1:} $y_1^* \leq y_2^*$.  In this case, Equation \ref{eqn:c} is equal to 0.  Moreover, in Equation \ref{eqn:b}, since $y'$ is restricted to be less than $y_2^*$, $F_{1|0}(y_1|y') < F_2(y_0|y')$ over this range; using this inequality and combining the terms from \Cref{eqn:a,eqn:b} implies
  \begin{align*}
    F_1^U(y_0, y_1) - F_2^U(y_0, y_1) &\geq \int_{\mathcal{Y}} (F_1(y_0|y') - F_2(y_0|y')) \indicator{ y' \geq y_1^*} \d{F_{t-1}(y')} \\
    &= F_2(y_0, y_1^*) - F_1(y_0, y_1^*) \geq 0
  \end{align*}
  where the last inequality holds because $F_1 \prec^{SI} F_2 \implies F_1 \prec^{C} F_2$ (\citet[Theorem 2.12]{joe-1997}) where $F \prec^{C} G$ is the concordance ordering and indicates that $F(x_1,x_2) \leq G(x_1,x_2)$ for all $x_1$ and $x_2$ and where $F$ and $G$ are joint distributions with the same marginals.

  \textit{Case 2:} $y_1^* > y_2^*$. In this case, Equation \ref{eqn:b} is equal to 0.  Also, Equation \ref{eqn:c} is greater than or equal to zero because $y'$ is restricted to be greater than $y_2^*$.  Equation \ref{eqn:a} is then given by
  \begin{align*}
    \int_{\mathcal{Y}} (F_1(y_0|y') - F_2(y_0|y')) \indicator{ y' \geq y_1^*} \d{F_{t-1}(y')} = F_2(y_0, y_1^*) - F_1(y_0, y_1^*) \geq 0
  \end{align*}
  where the first part holds because $y_1^* > y_2^*$ and where the last inequality holds because $F_1 \prec^{C} F_2$.  Therefore, in this case, all the terms in \Cref{eqn:a,eqn:b,eqn:c} are greater than or equal to zero and therefore the result holds.

  \bigskip
  
  \subproof{Lower Bound:} \mbox{}

  Let $F_1^L(y_0,y_1)$ denote the upper bound of the joint distribution given in Theorem \ref{theorem:joint} when $F_{Y_{0t}|Y_{0t-1},D=1} = F_1$ and likewise let $F_2^L(y_0, y_1)$ denote the upper bound when $F_{Y_{0t}|Y_{0t-1},D=1}=F_2$.  For this part, the goal is to show that $F_2^L(y_0,y_1) - F_1^L(y_0,y_1) \geq 0$ for all $(y_0, y_1) \in \mathcal{Y} \times \mathcal{Y}$.  Similar to the case for the upper bound, one can show that
  \begin{align}
    & F_2^L(y_0,y_1) - F_1^L(y_0,y_1) \nonumber \\
    & \hspace{1pt} = \int_{\mathcal{Y}} (F_2(y_0|y') - F_1(y_0|y')) \indicator{F_{1|0}(y_1|y') + F_1(y_0|y') - 1 \geq 0, F_{1|0}(y_1|y') + F_2(y_0|y') - 1 \geq 0} \ dF_{t-1}(y') \nonumber \\
    & \hspace{1pt} - \int_{\mathcal{Y}} (F_{1|0}(y_1|y') + F_1(y_0|y') - 1) \indicator{F_{1|0}(y_1|y') + F_1(y_0|y') - 1 \geq 0, F_{1|0}(y_1|y') + F_2(y_0|y') - 1 < 0} \ dF_{t-1}(y') \nonumber \\
    & \hspace{1pt} + \int_{\mathcal{Y}} (F_{1|0}(y_1|y') + F_2(y_0|y') - 1) \indicator{F_{1|0}(y_1|y') + F_1(y_0|y') - 1 < 0, F_{1|0}(y_1|y') + F_2(y_0|y') - 1 \geq 0} \ dF_{t-1}(y') \label{eqn:prop1b}
  \end{align}
  which holds by the same arguments as for the upper bound, but now using the conditional lower bounds on the joint distribution.  
  Because $F_{1|0}$, $F_1$, and $F_2$ are all stochastically increasing, there exist $y_1^\dagger = y_1^\dagger(y_0,y_1)$ and $y_2^\dagger = y_2^\dagger(y_0,y_1)$ such that
  \begin{align*}
    F_{1|0}(y_1|y_1^\dagger) + F_1(y_0|y_1^\dagger) = 1 \quad and \quad F_{1|0}(y_1|y_2^\dagger) + F_2(y_0|y_2^\dagger) = 1
  \end{align*}
  and for $y' \in \mathcal{Y}$ such that
  \begin{align*}
    \ y' > y_1^\dagger \implies F_{1|0}(y_1|y') + F_1(y_0|y') < 1, &\qquad \ y' \leq y_1^\dagger \implies F_{1|0}(y_1|y') + F_1(y_0|y') \geq 1
  \end{align*}
  and
  \begin{align*}
    \ y' > y_2^\dagger \implies F_{1|0}(y_1|y') + F_2(y_0|y') < 1, &\qquad \  y' \leq y_2^\dagger \implies F_{1|0}(y_1|y') + F_2(y_0|y') \geq 1
  \end{align*}
  Similarly to the proof for the upper bound, there may not be any values of $y'$ in $\mathcal{Y}$ that satisfy the strict inequalities above.  
  One can plug these into Equation \ref{eqn:prop1b} to obtain
  \begin{align}
    & F_2^L(y_0,y_1) - F_1^L(y_0,y_1) \nonumber \\
    & \hspace{50pt} = \int_{\mathcal{Y}} (F_2(y_0|y') - F_1(y_0|y')) \indicator{y' \leq y_1^\dagger, y' \leq y_2^\dagger} \ dF_{t-1}(y') \label{eqn:la}\\
    & \hspace{50pt} - \int_{\mathcal{Y}} (F_{1|0}(y_1|y') + F_1(y_0|y') - 1) \indicator{y_2^\dagger < y' \leq y_1^\dagger} \ dF_{t-1}(y') \label{eqn:lb}\\
    & \hspace{50pt} + \int_{\mathcal{Y}} (F_{1|0}(y_1|y') + F_2(y_0|y') - 1) \indicator{y_1^\dagger < y' \leq y_2^\dagger}  \ dF_{t-1}(y') \label{eqn:lc}
  \end{align}
  \textit{Case 1:} $y_2^\dagger \leq y_1^\dagger$

  In this case, Equation \ref{eqn:lc} is equal to 0.  For Equation \ref{eqn:lb}, $F_{1|0}(y_1|y') \leq 1 - F_2(y_0|y')$ when $y' \geq y_2^\dagger$ which implies that
  \begin{align*}
    & F_2^L(y_0,y_1) - F_1^L(y_0,y_1) \nonumber \\
    & \hspace{50pt} \geq \int_{\mathcal{Y}} (F_2(y_0|y') - F_1(y_0|y')) \indicator{y' \leq y_1^\dagger} \ dF_{t-1}(y')\\
    & \hspace{50pt} = F_2(y_0, y_1^\dagger) - F_1(y_0, y_1^\dagger) \geq 0
  \end{align*}
  which holds because $F_1 \prec^{C} F_2$ as implied by the assumptions in the proposition and which implies the result.

  \noindent \textit{Case 2:} $y_2^\dagger > y_1^\dagger$

  In this case, Equation \ref{eqn:lb} is equal to 0.  Equation \ref{eqn:lc} is greater than or equal to zero because $F_{1|0}(y_1|y') + F_2(y_0|y') > 1$ for $y' < y_2^\dagger$ (and equality holding when there is no $y'\in \mathcal{Y}$ that is strictly less than $y_2^\dagger$).  And Equation \ref{eqn:la} is equal to $F_2(y_0,y_1^\dagger) - F_1(y_0, y_1^\dagger)$ which holds because $y_1^\dagger < y_2^\dagger$ in this case), and this term is greater than or equal to zero because $F_1 \prec^C F_2$.  Thus, in this case, each term in \Cref{eqn:la,eqn:lb,eqn:lc} is greater than or equal to zero which implies the result.
  
  \pagebreak
  
  \section{Tables and Figures}

  \renewcommand{\arraystretch}{1}
\newcolumntype{.}{D{.}{.}{-1}}
\ctable[caption={Summary Statistics},label=tab:ss,pos=!h,]{lcccc}{\tnote[]{\textit{Notes:}  Summary statistics for individuals based on whether or not an individual was displaced from his job in 2008 or 2009.  The top panel uses the sample used for the main results in the paper with a sample size of 2,775 of which 122 are displaced which amounts to 5.6\% of the observations.  The bottom panel uses a balanced panel subset of the data for which earnings are available in all years in the table which includes 2,077 observations.  Earnings are in thousands of dollars. 

    \textit{Sources:} 1979 National Longitudinal Study of Youths}}{\FL
\multicolumn{1}{l}{}&\multicolumn{1}{c}{Displaced}&\multicolumn{1}{c}{Non-Displaced}&\multicolumn{1}{c}{Diff}&\multicolumn{1}{c}{P-val on Diff}\ML
{\bfseries Characteristics}&&&&\NN
~~Less than HS&0.10&0.08&0.02&0.43\NN
~~High School&0.65&0.61&0.04&0.41\NN
~~College&0.25&0.31&-0.06&0.19\NN
~~Hispanic&0.20&0.17&0.03&0.47\NN
~~Black&0.35&0.25&0.10&0.02\NN
~~White&0.45&0.57&-0.12&0.01\NN
~~Male&0.56&0.52&0.04&0.33\NN
~~Female&0.44&0.48&-0.04&0.33\ML
{\bfseries Path of Earnings}&&&&\NN
~~2013 Earnings&49.82&69.50&-19.68&0.01\NN
~~2011 Earnings&45.51&67.06&-21.55&0.00\NN
~~2009 Earnings&53.70&62.28&-8.57&0.17\NN
~~2007 Earnings&59.17&60.93&-1.76&0.78\NN
~~2005 Earnings&55.32&54.97&0.35&0.95\NN
~~2003 Earnings&49.71&49.88&-0.17&0.97\NN
~~2001 Earnings&42.71&46.58&-3.87&0.40\LL
}
\renewcommand{\arraystretch}{1}

\begin{figure}[t]
  \caption{Bounds on the Quantile of the Treatment Effect}
  \begin{subfigure}{.5\textwidth}
    { \centering \includegraphics[width=.95\textwidth]{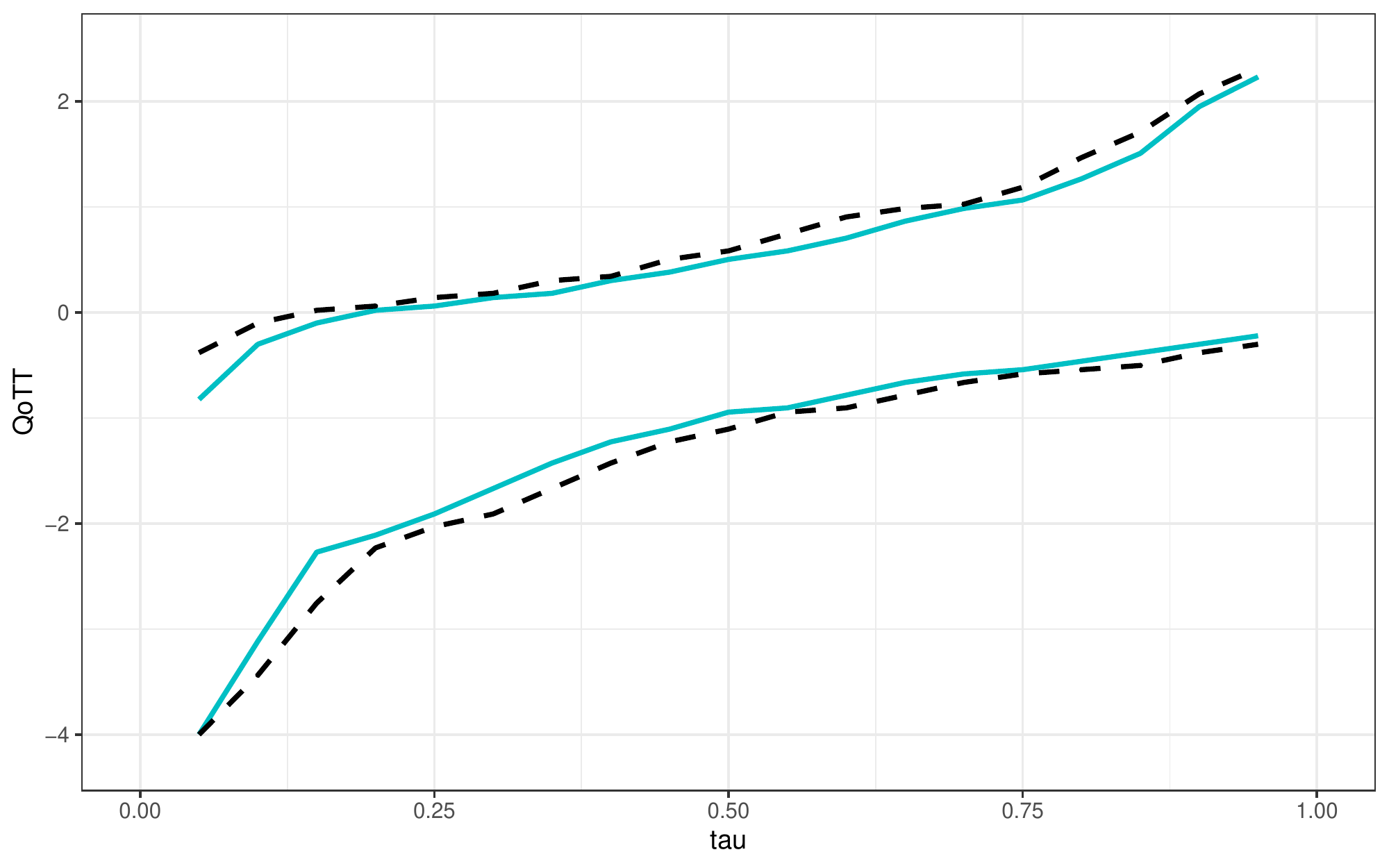} \\ }
    \caption{Worst Case Bounds}
    \label{fig:wcb}
  \end{subfigure}
  \begin{subfigure}{.5\textwidth}
    { \centering \includegraphics[width=.95\textwidth]{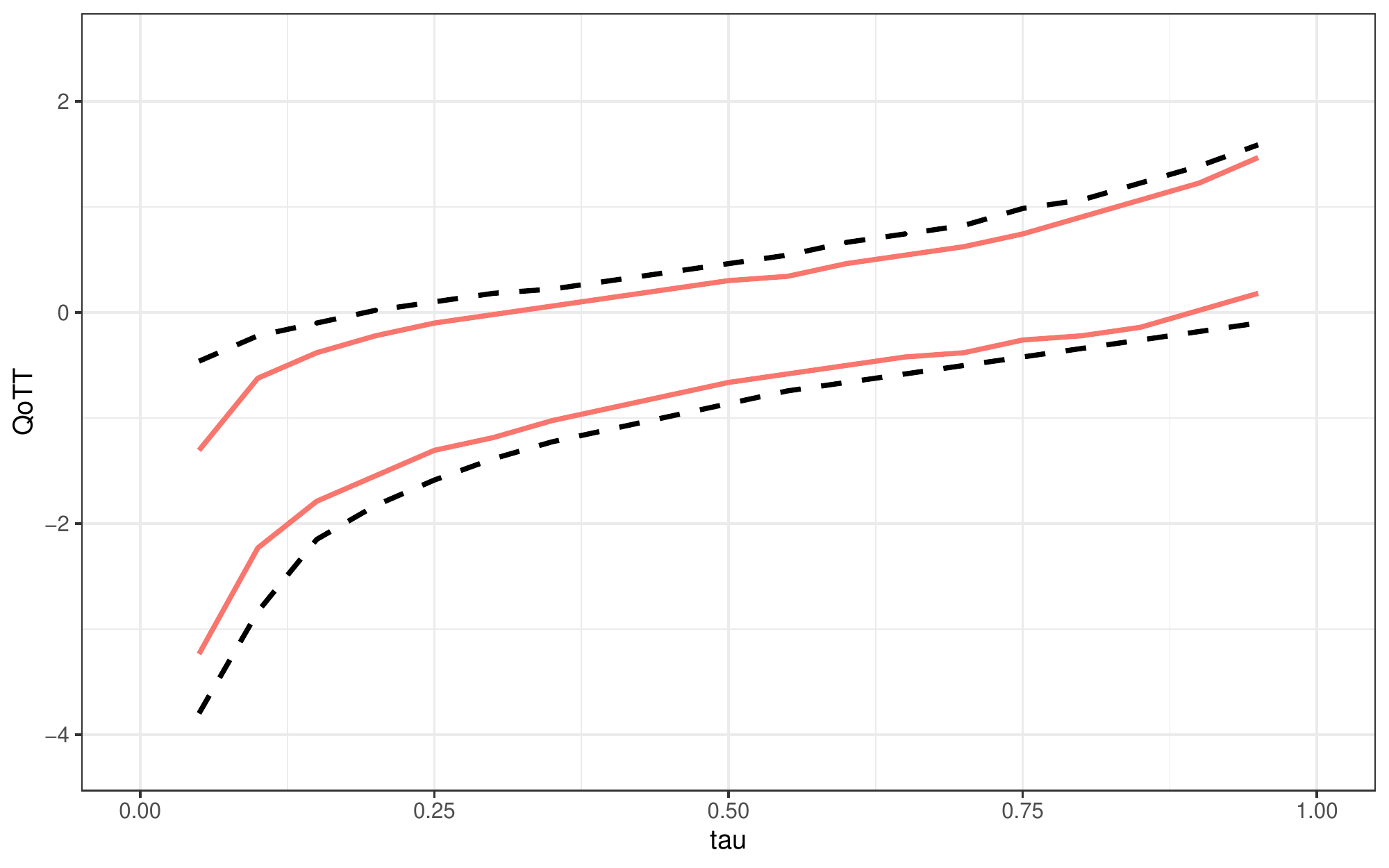} \\ }
    \caption{Bounds under Copula Stability Assumption}
    \label{fig:csab}
  \end{subfigure}
  { \footnotesize \textit{Notes:}  The figure provides bounds on the Quantile of the Treatment Effect on the Treated.  Panel (a) contains worst-case bounds that only use information from the marginal distributions of treated and untreated potential outcomes.  Panel (b) contains bounds that come from using the method developed in the current paper under the \ref{ass:copula-stability}.  In each panel, the scale of the y-axis is in log points.  Most of the reported results in the text convert log points into percentage changes (see Footnote \ref{fn:pc}).  The dotted lines provide 95\% confidence intervals for the estimated lower and upper bounds using the numerical bootstrap as discussed in the text.

  \textit{Sources:}  1979 National Longitudinal Survey of Youth}
\end{figure}

\begin{figure}[t]
  \caption{Comparison of Bounds under Different Conditions}
  \begin{center}
    \begin{subfigure}{.35\linewidth}
      \begin{center}
        \includegraphics[width=.95\linewidth]{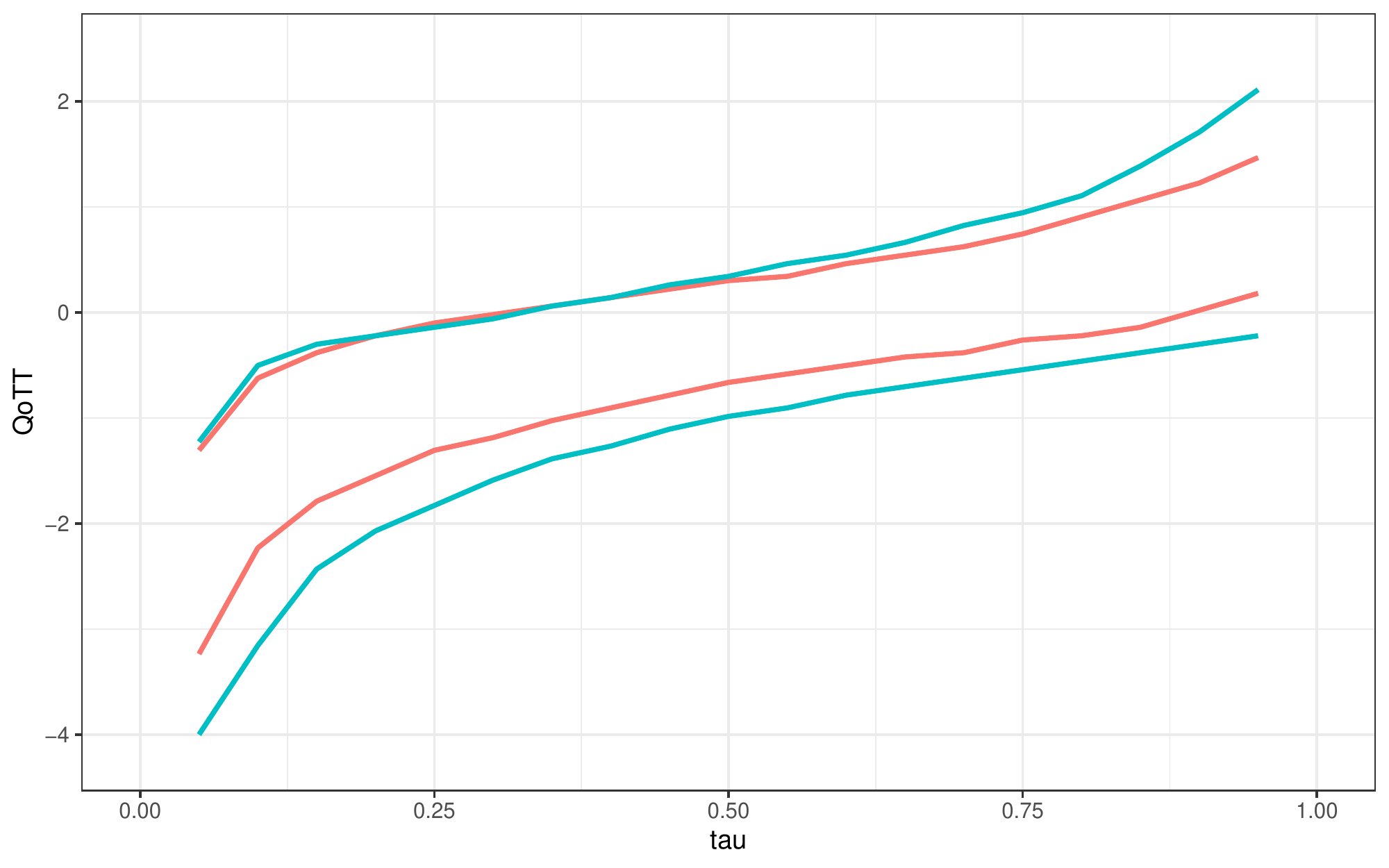} 
        \caption{CSA $+$ WC}
      \end{center}
  \end{subfigure}
  \begin{subfigure}{.35\linewidth}
    \begin{center}
      \includegraphics[width=.95\linewidth]{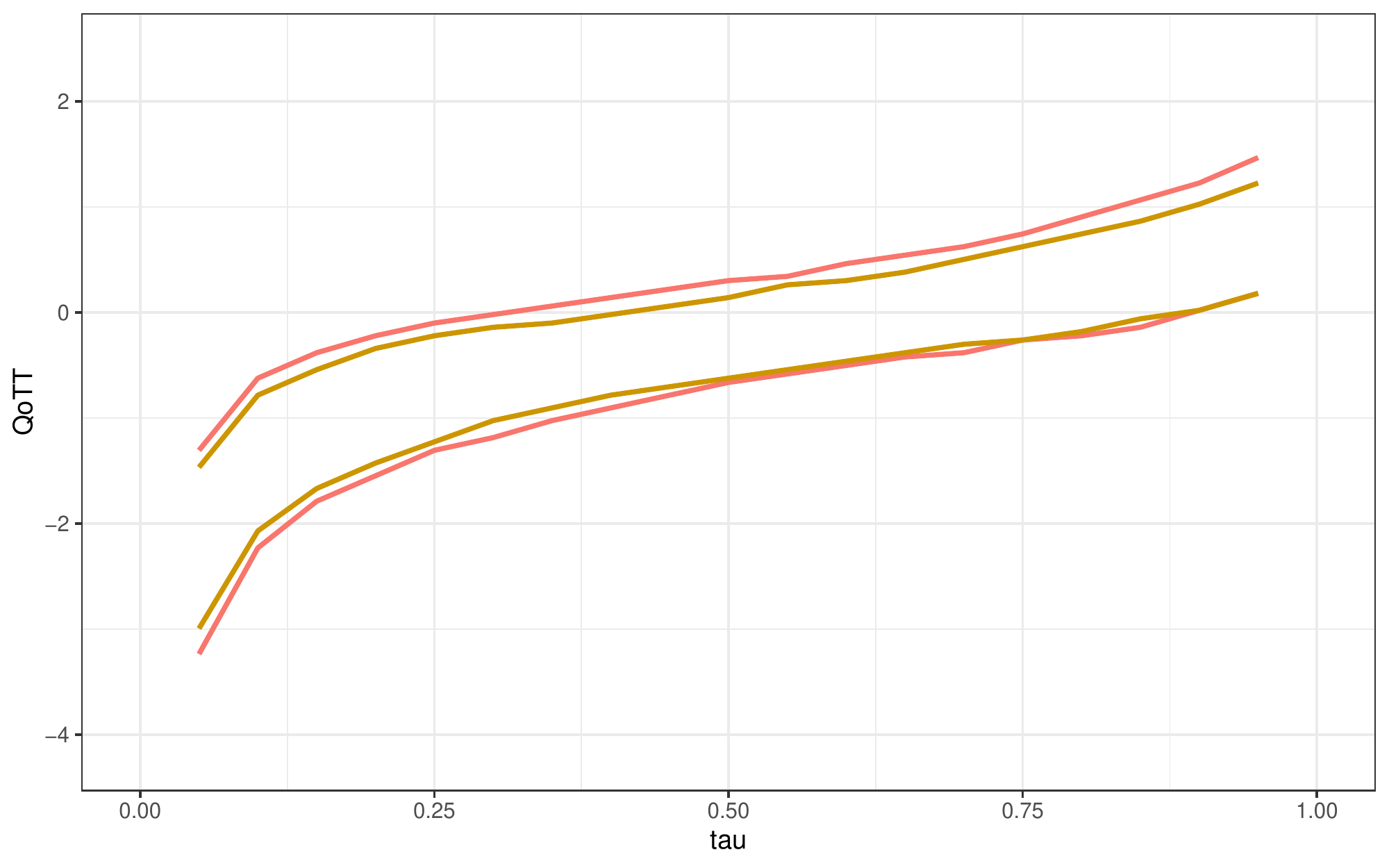}
      \caption{CSA $+$ CSA/Covs}
    \end{center}
  \end{subfigure}
  
  \begin{subfigure}{.35\linewidth}
    \begin{center}
      \includegraphics[width=.95\linewidth]{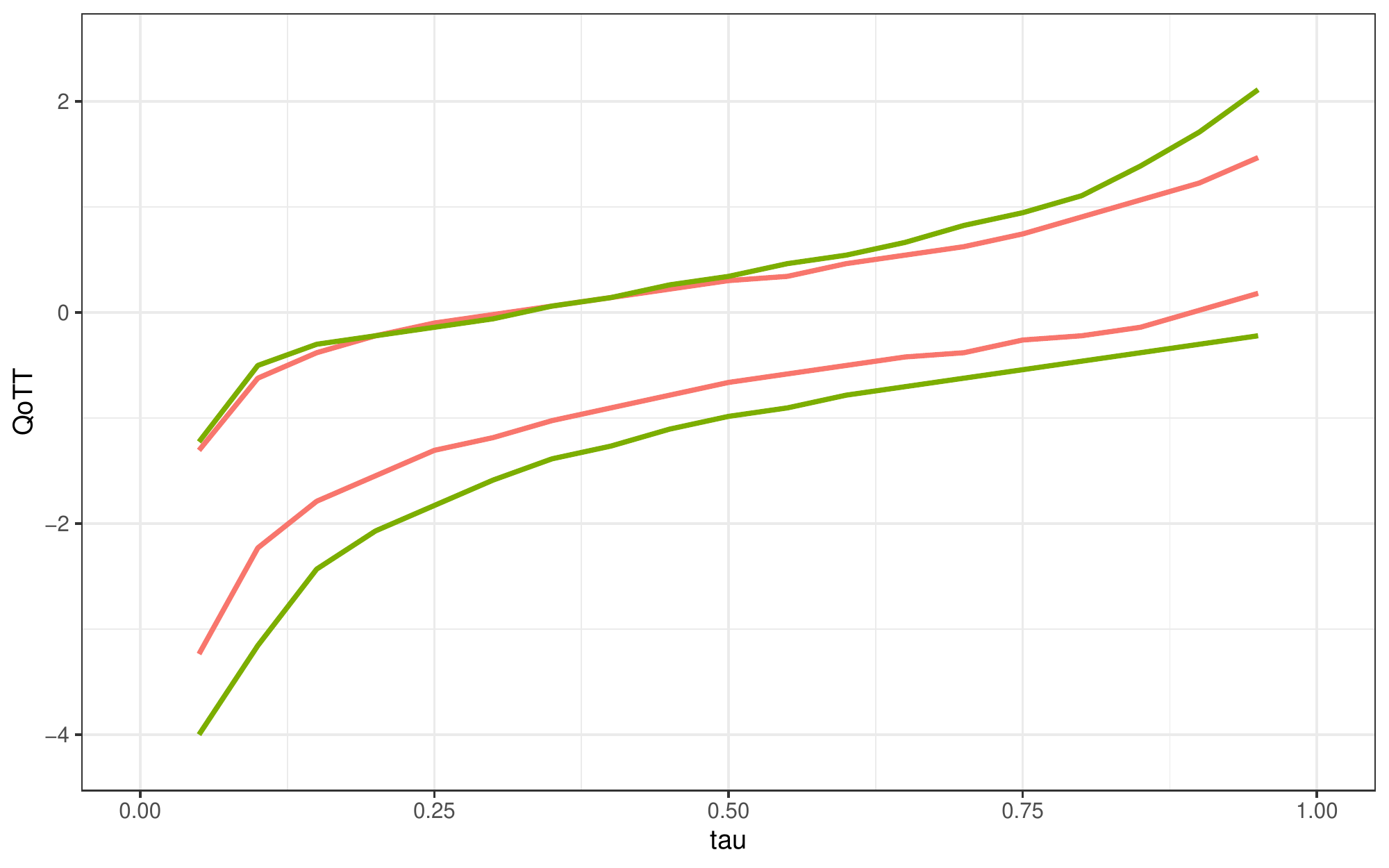} 
      \caption{CSA $+$ Covs}
    \end{center}
  \end{subfigure}
  \begin{subfigure}{.35\linewidth}
    \begin{center}
      \includegraphics[width=.95\linewidth]{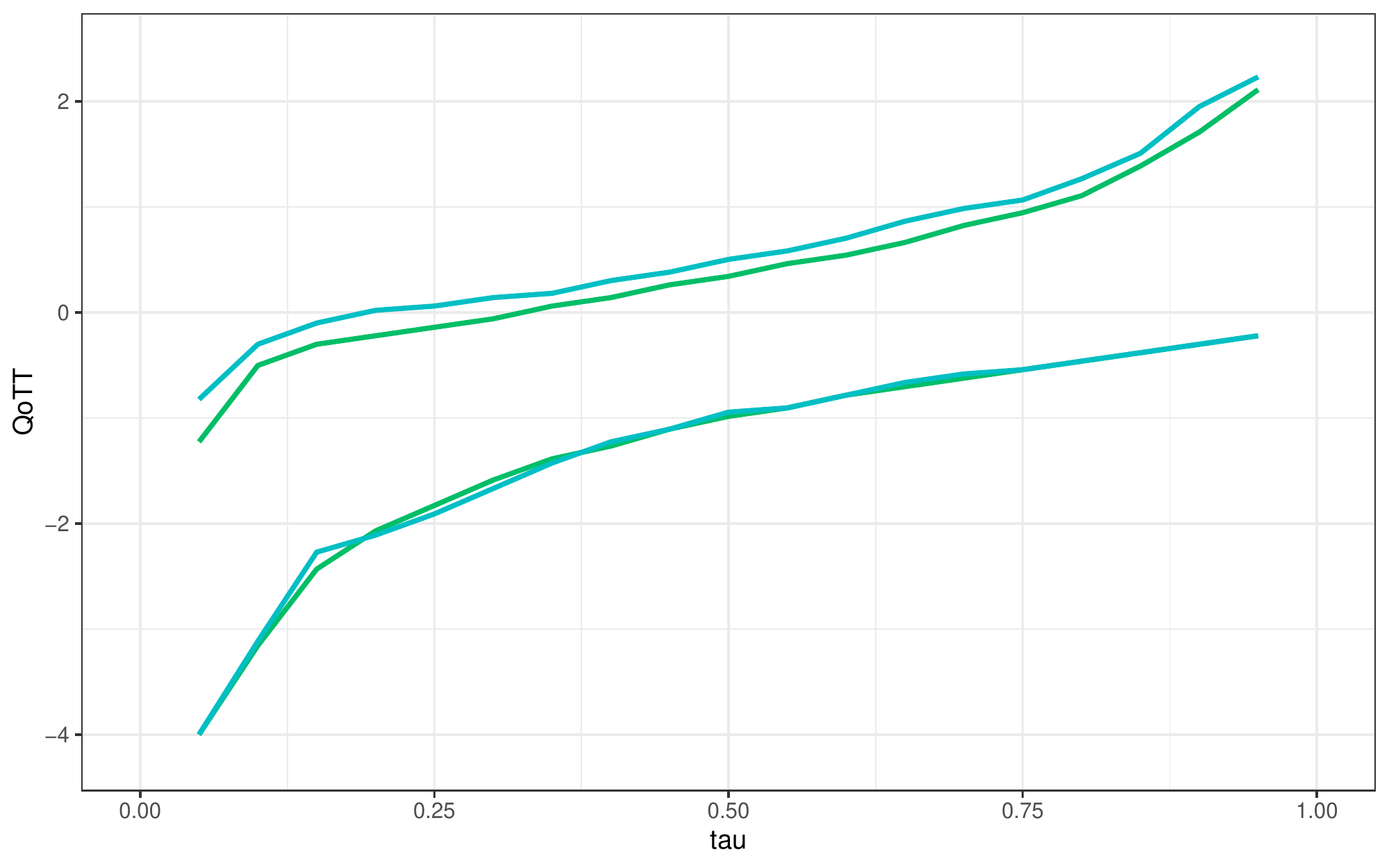}
      \caption{WC $+$ Covs}
    \end{center}
  \end{subfigure}
  \end{center}

  { \footnotesize \textit{Notes:}  The figure includes bounds under different conditions presented in the main text.  The top left panel includes bounds under the \ref{ass:copula-stability} along with the worst-case bounds (these are the same as what is reported in \Cref{fig:wcb,fig:csab}).  The top right panel compares bounds under the \ref{ass:copula-stability} in the unconditional case to bounds under \ref{ass:copula-stability} that also use covariates to further tighten the bounds.  The bottom left panel compares bounds under the \ref{ass:copula-stability} to bounds that tighten the worst-case bounds only using covariates.  The bottom right panel compares the worst-case bounds to the bounds that only use covariates to obtain tighter bounds. The scale of the y-axis is in log points.  Most of the reported results in the text convert log points into percentage changes (see Footnote \ref{fn:pc}).  Plots that include confidence intervals for the bounds in each case are available in the Supplementary Appendix.  %
  
  \textit{Sources:}  1979 National Longitudinal Survey of Youth}
  \label{fig:compare-bounds}
\end{figure}

\begin{figure}[t]
  \caption{Spearman's Rho for Every Four Years 1987-2011}
  { \centering \includegraphics[width=.6\linewidth]{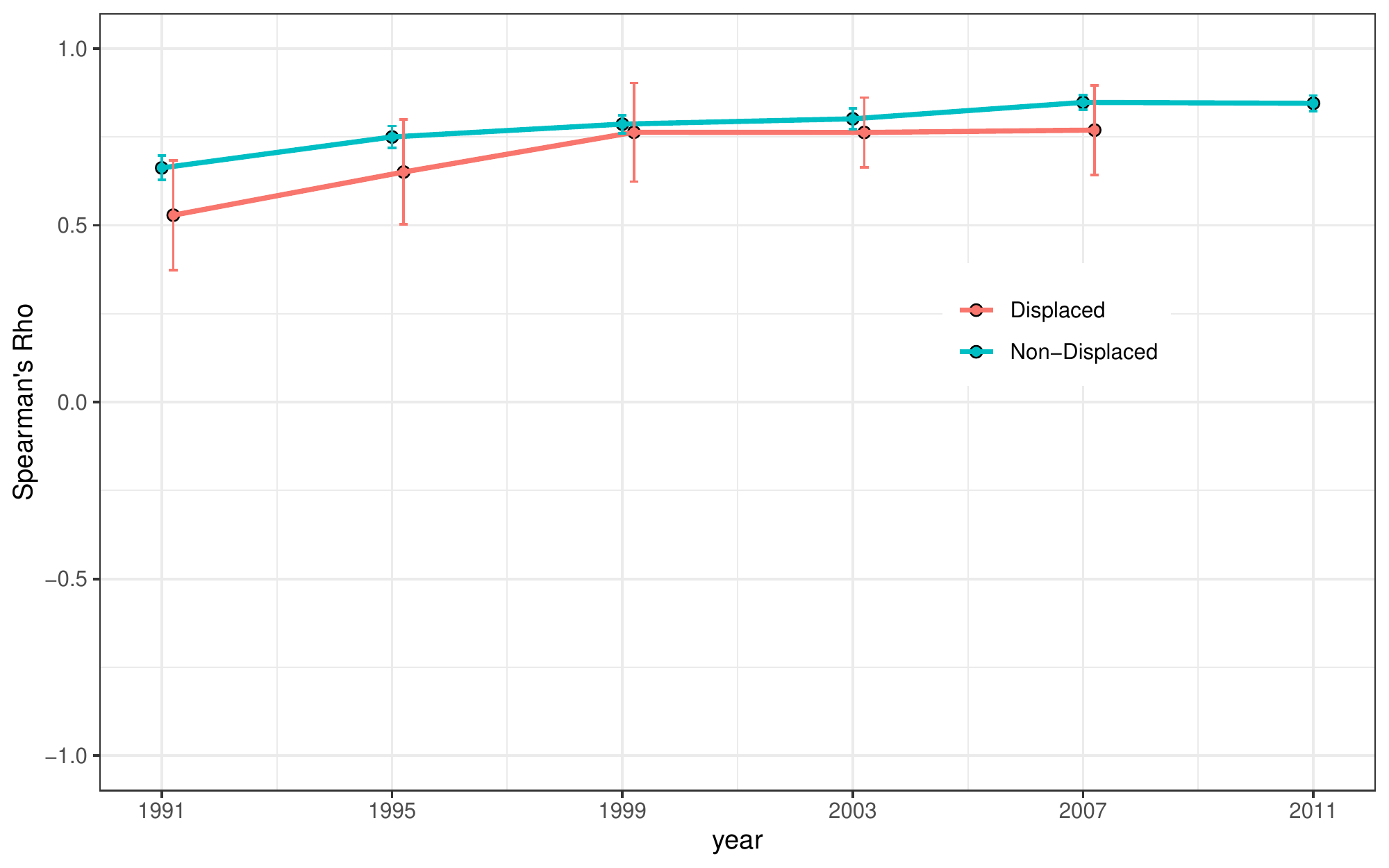} \\ }
  \label{fig:spearmans-rho}

  { \footnotesize \textit{Notes: }  This figure provides estimates of Spearman's Rho for the group of displaced workers and the group of non-displaced workers.  Spearman's Rho is the correlation of the ranks of earnings in period $s$ and $s-1$ and depends only on the copula of earnings in period $s$ and period $s-1$.  The sample includes a subset of the dataset used in the main analysis that includes 1,993 individuals that have positive earnings in each year from 1987-2011.  Standard errors are computed using the block bootstrap with 1000 iterations.

  \textit{Sources: }  1979 National Longitudinal Survey of Youth}
\end{figure}

\end{document}